\pdfoutput=1
\documentclass[sigconf]{acmart}
\renewcommand\footnotetextcopyrightpermission[1]{} % removes footnote with conference information in first column
\usepackage{booktabs}
\setcopyright{rightsretained}
\usepackage{multirow} 
\usepackage{flushend}

\usepackage{graphicx}
\usepackage[utf8]{inputenc}

\usepackage{amssymb}
\usepackage{url}
\usepackage{hyperref}
\usepackage{wrapfig}
\usepackage{units}

\usepackage[printonlyused]{acronym}

\usepackage{calc}
\usepackage{amsmath}
\usepackage{amssymb}
\usepackage{amsfonts}
\usepackage{mathtools}
\usepackage{color, colortbl}
\usepackage{multirow} % to merge multiple rows
\usepackage[caption=false]{subfig}
\usepackage{stfloats}

\usepackage{cleveref}
\usepackage{listings}
\usepackage{tikz}
\usetikzlibrary{arrows,positioning,automata,shadows,fit,shapes}

\definecolor{mygreen}{RGB}{28,172,0} % color values Red, Green, Blue
\definecolor{mylilas}{RGB}{170,55,241}

\definecolor{BgGray}{gray}{0.7}%
\definecolor{BgGray2}{gray}{0.96}%
\definecolor{RowColorOdd}{named}{BgGray2}%
\definecolor{RowColorEven}{named}{white}%
\definecolor{comments}{gray}{.5}
\definecolor{Gray}{gray}{0.85}

\usepackage{pifont}% http://ctan.org/pkg/pifont

\newcommand{\lte}{l}
\newcommand{\wifi}{w}
\newcommand{\user}{u}
\newcommand{\distance}{d}
\newcommand{\userAngle}{\theta}
\newcommand{\numLTEantenna}{K}
\newcommand{\distanceWiFiLTE}{D}  
\newcommand{\usersSetLTE}{\mathcal{U}^{\lte}}
\newcommand{\usersSetWiFi}{\mathcal{U}^{\wifi}}
\newcommand{\userLTE}{\user^{\lte}}
\newcommand{\userWiFi}{\user^{\wifi}}
\newcommand{\numUsersLTE}{M}
\newcommand{\numUsersWiFi}{N}
\newcommand{\priority}{\beta}
\newcommand{\rate}{r}
\newcommand{\throughput}{R}
\newcommand{\AntennaConfig}{\Phi}
\newcommand{\nullPrefix}{\varnothing}
\newcommand{\airtime}{\alpha}
\newcommand{\delay}{\tau}
\newcommand{\pathlossCoef}{\gamma}
\newcommand{\bandwidth}{B}

\newcommand{\threshToDetectLTE}{\Gamma_{l}}
\newcommand{\threshToDetectWiFi}{\Gamma_{w}}
\newcommand{\csRangeFlag}{\sigma}

\newcommand{\heuristicname}{GREEDY}
\newcommand{\nonull}{NoNull}
\newcommand{\optname}{OptMaxSum}

\newcommand{\maxWiFigainthru}{44} % OPTIMAL achieves 40 percent for a single user {8}
\newcommand{\maxLTEgainthru}{221}

\usepackage{etoolbox}
\makeatletter
\patchcmd{\maketitle}{\@copyrightspace}{}{}{}
\makeatother
\begin{document}

\title[Coexistence Gaps in Space]{Coexistence Gaps in Space: Cross-Technology Interference-Nulling for Improving LTE-U/WiFi Coexistence}

%\renewcommand{\shorttitle}{An Ultra-Wide Overlay Cognitive Radio System for Wireless Backhauling ...}
%\titlenote{Produces the permission block, and copyright information}
%\subtitle{Extended Abstract}
%\subtitlenote{The full version of the author's guide is available as \texttt{acmart.pdf} document}

\author{Suzan Bayhan, Anatolij Zubow, and Adam Wolisz}
\affiliation{%
  \institution{Technische Universität Berlin}
}
\email{{bayhan,zubow,wolisz}@tkn.tu-berlin.de}

%% Abstract
\begin{abstract}

To avoid the foreseeable spectrum crunch, LTE operators have started to explore the option to directly use 5\,GHz unlicensed spectrum band being used by IEEE 802.11~(WiFi).
However, as LTE is not designed with shared spectrum access in mind, there is a major issue of coexistence with WiFi networks.
Current coexistence schemes to be deployed at the LTE-U BS create coexistence gaps only in one domain~(e.g., time, frequency, or space) and can provide only incremental gains due to the lack of coordination among the coexisting WiFi and LTE-U networks.
Therefore, we propose a coordinated coexistence scheme which relies on cooperation between neighboring LTE-U and WiFi networks.
Our proposal suggests that LTE-U BSs equipped with multiple antennas can create coexistence gaps 
in space domain in addition to the time domain gaps by means of cross-technology interference nulling towards WiFi nodes in the interference range. In return, LTE-U can increase its own airtime utilization while trading off slightly its antenna diversity.
The cooperation offers benefits to both LTE-U and WiFi in terms of improved throughput and decreased channel access delay.
More specifically, system-level simulations reveal a throughput gain up to \maxLTEgainthru\% for LTE-U network and \maxWiFigainthru\% for WiFi network depending on the setting, e.g., distance between the two cell, number of LTE antennas, and WiFi users in the LTE-U BS neighborhood.
Our proposal provides significant benefits especially for moderate separation distances between LTE-U/WiFi cells where interference from a neighboring network might be severe due to the hidden network problem.

\end{abstract}

\keywords{LTE unlicensed, LTE-U, WiFi, co-existence, interference nulling}

\maketitle
% personal use notification
%\thispagestyle{fancy}

%% Introduction
%!TEX root = main.tex
\section{Introduction}\label{sec:introduction}

The rapid growth of wireless traffic, known as data tsunami, has been a key challenge for mobile network operators in the past few years.
Luckily, wireless local area networks~(WiFi/IEEE 802.11) have acted as a life ring by carrying a significant fraction of the offloaded mobile data traffic~(60\% in 2015~\cite{cisco_march2017}) which would otherwise follow the cellular network.
However, LTE operators have started to explore other options, known as \textit{unlicensed LTE}, to use the unlicensed spectrum directly by performing carrier aggregation deep at the radio link level, i.e. modem-level. 
This gives better load balancing on the licensed and unlicensed channels as the LTE network is always aware of the network load and signal quality of both the licensed and unlicensed links and can balance traffic on the links accordingly~\cite{lteu_whitepaper_qualcom15}. 
While such an approach has potential to expand the cellular capacity significantly, it is expected that the proliferation of a particular technology like LTE in Unlicensed spectrum (LTE-U~\cite{lteu_forum}) combined with the expected exponential growth in the usage of WiFi will result in severe mutual interference.
Therefore, both networks operating on the same channel at 5 GHz UNII bands will experience a significant performance degradation, e.g., \cite{jindral2015,olbrich2017wiplus}, unless LTE-U networks implement coexistence solutions cautiously.

\begin{figure}[h!]
	\centering
	\includegraphics[width=0.8\linewidth]{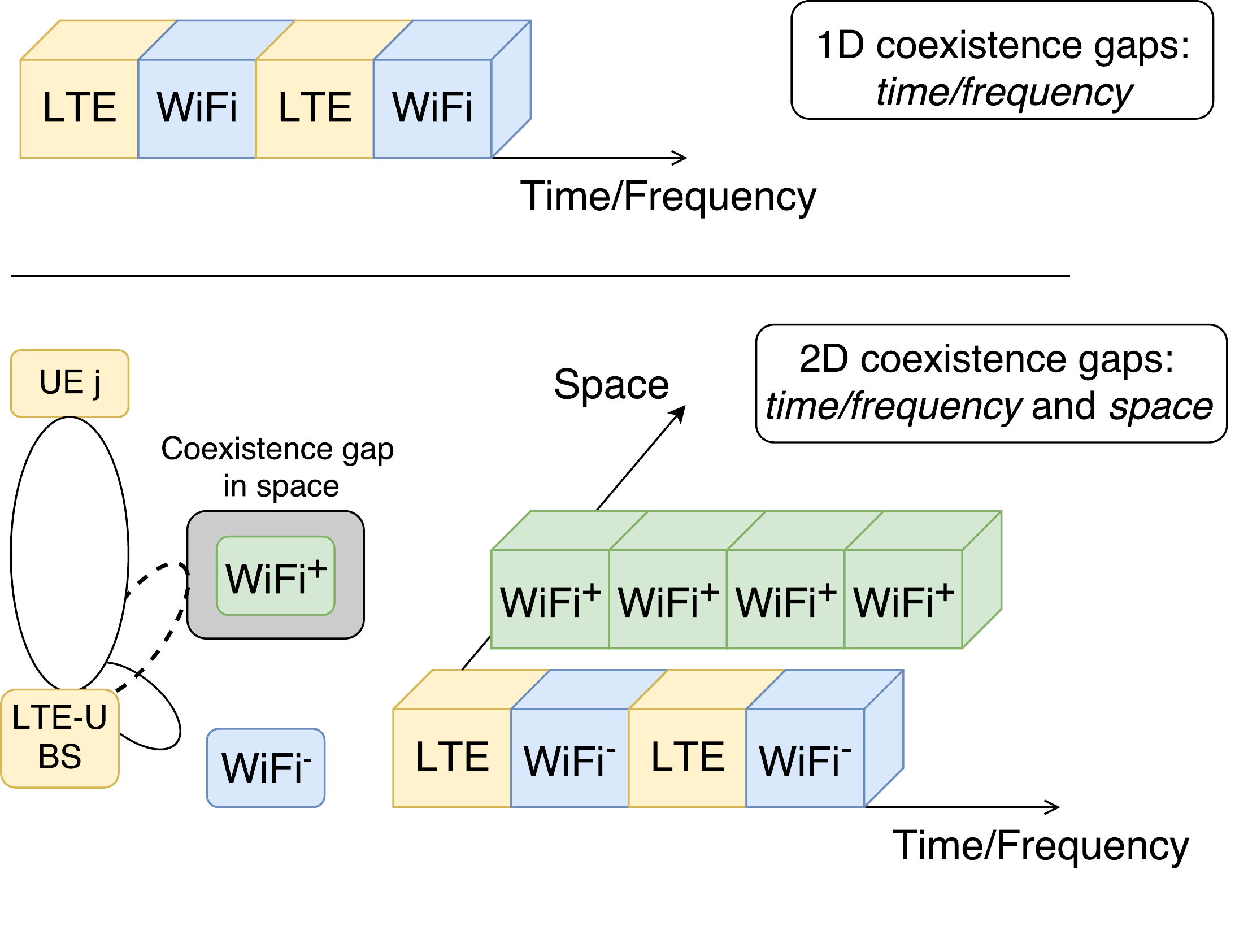}
	\vspace{-10pt}
	\caption{Coexistence gaps in one dimension either in time or frequency domain~(top figure) and proposed coexistence gaps in two dimensions~(bottom figure).
	\textrm{WiFi}$^+$ is the WiFi node being nulled and \textrm{WiFi}$^-$ receives strong signals from the LTE BS as it is not nulled.}
	\vspace{-5pt}
	\label{fig:coexgaps}
\end{figure}

Recent years have witnessed a boom of noncoordinated coexistence designs for LTE-U and WiFi, e.g., \cite{beluri2012mechanisms,almeida2013enabling,zhang2015coexistence,al20155g,chaves2013lte}.
Such schemes are simple to realize as they do not require any underlying infrastructure that connects the systems sharing the band for the purpose of information exchange.
The majority of proposals focus only on the LTE-U network and target improving its coexistence-friendliness towards WiFi by adapting its operation parameters, e.g., duty-cycle and subframe puncturing, at the expense of performance in the LTE-U network.
However, such noncoordinated coexistence solutions can only provide incremental gains as the major bottleneck of the coexistence setting is overlooked: \textit{the lack of flexibility} of the LTE-U network due to its scheduled medium access without performing listen-before-talk~(LBT) making it hard to adapt in short-term to the coexistence setting.
We argue that flexibility of both coexisting networks is key to achieving "joy of the commons" as opposed to well-known issue of "tragedy of the commons" in the unlicensed bands where the networks operate mostly on equal rights.\footnote{Some technologies such as radar at 5 GHz unlicensed bands may be prioritized resulting in different priorities. 
Flexibility in other spectrum authorization regimes, e.g., licensed access, is not as crucial since the rules of sharing are mostly determined via contracts between the coexisting networks.}

WiFi is coexistence-friendly as it has high flexibility in the time domain, i.e., it uses the channel in a random access LBT manner with a fine time granularity.
On the contrary, the LTE-U network due to its missing LBT mechanism can only adapt to the dynamics of the coexistence setting in a longer time scale, e.g., in the order of tens of milliseconds.  
That means, LTE-U lacks flexibility in the time domain.
However, we can add flexibility to the LTE-U in the space domain which can mitigate its time domain inflexibility as shown in Fig.~\ref{fig:coexgaps}.
More specifically, we suggest that LTE-U BSs equipped with an antenna array, e.g.  Uniform Linear Array~(ULA), should exploit some of its antenna resources to decrease its impact from the Downlink~(DL) traffic in unlicensed channel to co-located WiFi nodes by performing interference-nulling towards them.
In return, LTE-U can increase its own airtime utilization, i.e. duty cycle, as the nulled WiFi nodes can receive their DL traffic during LTE-U's ON period without distortion and hence need not to be considered in airtime fairness considerations~(as will be explained in Sec~\ref{sec:airtime}).
In other words, an LTE-U BS can create coexistence gaps (similar to \textit{almost blank subframes}) in space domain by means of interference nulling.
We argue that our proposed scheme smoothly introduces politeness to the LTE-U  which is crucial for fair spectrum sharing with WiFi, instead of changing the LTE's nature to introduce LBT functionality.

Our proposal is beyond coexistence: it suggests direct cooperation among WiFi and LTE-U networks, which we believe is necessary for using the unlicensed bands with high efficiency rather than passively implementing coexistence solutions to decrease the impact of one network on the other.
Hence, it falls into the family of coordinated coexistence solutions~\cite{al20155g} that rely on interworked infrastructure to allow central control of spectrum access and/or sharing information between the systems.
Although current technology does not support communication of control messages between LTE-U and WiFi, a recent study, LtFi~\cite{gawlowicz2017ltfi}, shows that it is possible to create a cross technology control channel~(CTC) between LTE-U and WiFi for the purpose of radio resource management. 
We believe that our approach can be implemented using such a CTC.

\smallskip

\noindent \textbf{Contributions:} We extend the coexistence capability of  LTE-U networks by introducing coexistence gaps in space domain in addition to coexistence gaps in time domain.
More clearly, we propose to apply interference-nulling from LTE-U BSs equipped with multiple antennas towards co-located WiFi nodes using the same unlicensed channel as a way to create coexistence gaps in space so that coexistence between LTE-U and WiFi can be improved. 
We first present a model capturing the trade-offs between the airtime and the channel rate under a given nulling configuration.
%
%\tolja{Which forces???}\su{I think this was your sentence :) } 
% 
Next, we provide an optimization problem formulation to derive the optimal nulling configuration and also present a low-complexity heuristic for finding groups of nodes to be nulled. 
Simulation results reveal that interference-nulling can improve the throughput of the LTE-U cell up to  \maxLTEgainthru\% while also providing some gains for the WiFi, e.g., \maxWiFigainthru\%. 
Moreover, both systems enjoy lower channel access delay which is of great importance for applications requiring low-latency communication.

%
%Main idea is to use some of the LTE beams for nulling a selected set of WiFi stations such that their downlink traffic will only be slightly affected by a simultaneous LTE signal.
%
%In return, LTE SBS achieves longer airtime due to its on CSAT period adaptation philosophy which considers the coexisting WiFi nodes for fairness reasons. This increase in airtime is the key incentive for LTE to sacrifice some of its antenna resources for WiFi nulling.
%
%As nulling decision is by no means trivial, e.g., does the increased airtime outweigh the benefit from antenna diversity?, we devise several schemes to select which stations and how many of them to null.

% paper organization
The rest of the paper is organized as follows. 
Section~\ref{sec:lte_primer} provides a very brief overview of the necessary background on LTE-U, WiFi, and interference nulling which our work relies on. 
Section~\ref{sec:sysmodel} introduces the considered system model while
Section~\ref{sec:proposal} presents our proposal, i.e., optimal interference nulling for WiFi and LTE-U coexistence.
Section~\ref{sec:greedy} provides a low-complexity yet feasible solution to select which WiFi nodes to be nulled.
Section~\ref{sec:eval} assesses the performance of our proposal first by comparing it with the conventional approach where coexistence gaps are only in the time-domain, i.e., there is no interference nulling, and next by evaluating the impact of number of WiFi users, LTE-U BS antennas as well as the separation distance between LTE-U BS and the WiFi AP.
This section also discusses the limitations of our work.
Finally, Section~\ref{sec:conc} concludes the paper with a list of future work.

%% Primer

\section{Background} \label{sec:lte_primer}
We provide an overview of the LTE-U and WiFi standards as well as interference-nulling, which are relevant to our discussion.
\subsection{LTE-U}

LTE-U is being specified by the LTE-U forum~\cite{lteu_forum} as the first cellular solution using unlicensed bands for the downlink~(DL) traffic. 
The LTE carrier aggregation framework supports utilization of the unlicensed band as a secondary cell in addition to the licensed anchor serving as the primary cell~\cite{ni_lte_whitepaper}. 
The LTE-U channel bandwidth is set to 20\,MHz which corresponds to the smallest channel width in WiFi. 
The main coexistence mechanism of LTE-U is dynamic channel selection where the LTE-U BS seeks for a clear channel~(coexistence gap in frequency domain in Fig.~\ref{fig:coexgaps}). 
If no such channel is identified, the channel with the least observed WiFi channel utilization is selected and LTE-U applies duty-cycling for sharing the medium in the time domain.
As LTE-U does not implement LBT, it can be deployed in countries such as USA, China and India, where LBT is not required for unlicensed channel access.

\begin{figure}[h!]
	\centering
	\includegraphics[width=0.8\linewidth]{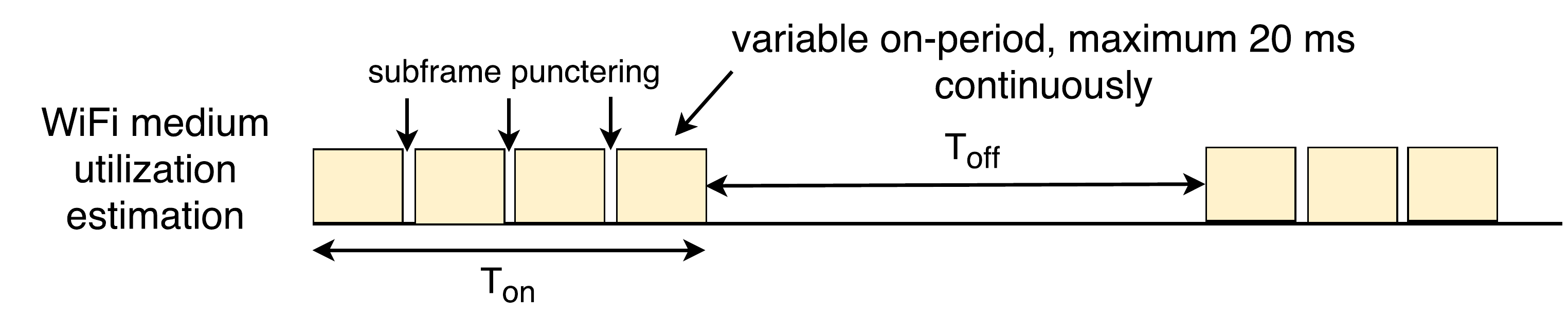}
	\vspace{-10pt}
	\caption{Adaptive duty cycling in LTE-U.}
	\vspace{-10pt}
	\label{fig:lte-u-overview}
\end{figure}

Fig.~\ref{fig:lte-u-overview} illustrates the duty cycled unlicensed channel access of LTE-U. 
LTE-U BSs actively observe the channel for WiFi transmissions to estimate channel activity for dynamic channel selection and adaptive duty cycling. 
A mechanism called \textit{carrier sense adaptive transmission}~(CSAT) is used to adapt the duty cycle~\cite{QualcommLTEPres15,lteu_qualcom2016}, i.e., by modifying the $T_{\mathrm{on}}$ and $T_{\mathrm{off}}$ values, to achieve fair sharing. 
Moreover, LTE-U transmissions contain frequent gaps in the on-period, which allow WiFi to transmit delay-sensitive data. 
Qualcomm~\cite{lteu_qualcom2016} recommends that LTE-U should use period of 40, 80 or 160\,ms and at least 2\,ms puncturing has to be applied every 20\,ms. 
Note that LBT is not applied in LTE-U before transmission of packets in the on-period.

\subsection{WiFi}

In contrast to LTE-U which uses scheduled channel access, WiFi nodes~(APs as well as STAs) perform random channel access using an LBT scheme, i.e. CSMA. 
WiFi makes use of both virtual and physical carrier sensing. 
Because WiFi is unable to decode LTE-U packets, it has to rely on physical carrier sensing~(CS). 
Moreover, CS is restricted to Energy Detection~(ED) which is less sensitive as compared to preamble-based CS methods: ED threshold for sensing an LTE-U signal is -62\,dBm whereas a WiFi AP can detect other WiFi signals at the sensitivity level around -82 dBm.\footnote{There is also an ongoing debate on whether WiFi's ED threshold is fair. 3GPP has requested to increase WiFi's ED level from -62\,dBm to -72\,dBm~\cite{IEEEMentorEDThreshPres2017}.}
ED threshold for LTE-U to sense WiFi signals is -82\,dBm which is recently agreed by the WiFi Alliance's Coexistence Test Plan~\cite{WFACoexTestPlan}. 

An LTE-U's transmission may have the following two impacts on WiFi depending on the received LTE-U signal's strength:
(i) WiFi cannot access the medium during LTE-U's on-periods as ED mechanism of WiFi is triggered at the WiFi transmitter; (ii) WiFi experiences frequent packet corruptions due to co-channel interference at the WiFi receiver.
Case~(i) results in lower available airtime for WiFi due to channel contention while Case (ii) results in wasted airtime due to packet loss caused by inter-technology hidden node problem~\cite{jindral2015,olbrich2017wiplus}.

\subsection{Interference Nulling}\label{sec:nulling}

A transmitter equipped with an antenna array, e.g. ULA, can use precoding to change how its signal is received at a particular wireless node. To do so, it multiplies the transmitted signal by a precoding matrix $P$. Specifically, in interference nulling the precoding matrix is chosen to null~(i.e., cancel) the signal at a particular receiver, i.e. $HP = 0$, where $H$ is the channel matrix from transmitter to receiver~\cite{lin2011random}. Note that the transmitter requires knowledge of $H$.

%% system model

\section{System Model}\label{sec:sysmodel}

We consider a coexistence scenario where an LTE-U cell and WiFi Basic Service Set~(BSS) have overlapping coverage and share the same unlicensed channel for their operation.~\footnote{
An extension to multiple cells is straightforward in case LTE cells are synchronized in their CSAT cycles. We plan to extend to a more generic setting in our future work.}
 \begin{figure}[h!]
	\centering
	\includegraphics[scale=0.30]{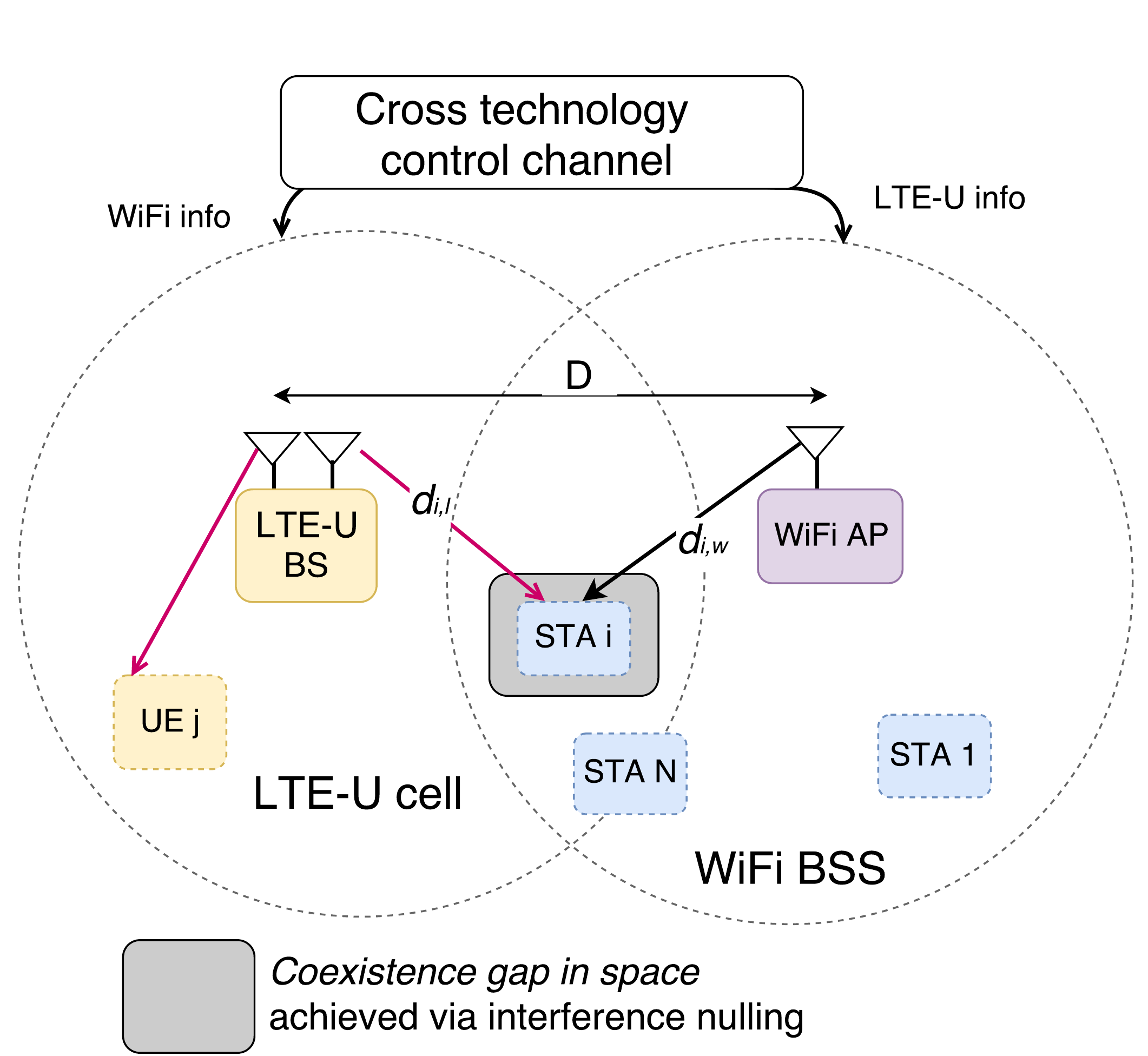}
	\caption{Considered coordinated LTE-U and WiFi coexistence setting. 
	}
	\vspace{-5pt}
	\label{fig:sysmodel}
\end{figure}
Consider a system as in Fig.~\ref{fig:sysmodel} where there is an LTE-U BS and a WiFi AP, separated by distance \distanceWiFiLTE.
Denote the set of UEs served by the LTE-U BS by $\usersSetLTE=\{ \userLTE_1, \cdots, \userLTE_\numUsersLTE\}$.
Similarly, denote the set of stations served by the WiFi AP by $\usersSetWiFi=\{ \userWiFi_1, \cdots, \userWiFi_\numUsersWiFi\}$.
For the simplicity of the notations, we will use index 0 to denote the WiFi AP~($\userWiFi_{0}$) and LTE-U BS~($\userLTE_0$).
Let $\distance_{i,x}$ and $\userAngle_{i,x}$ denote the distance and angle of a user $i$~(be it a UE or STA) from a BS $x$~($x=\lte$ for LTE-U or $\wifi$ for WiFi AP), respectively.
We assume that LTE-U BS serves its UEs in different time slots, i.e. TDMA based scheduling. 

We consider backlogged traffic for both networks and focus on the DL only.
For LTE-U system, this corresponds to supplementary DL case.
For WiFi, our scenario is still relevant as current networks are DL-heavy, e.g., 80-90\% of data traffic is attributed to DL~\cite{ITU_traffic_est_report}. 

LTE-U BS detects the WiFi nodes if it receives the WiFi signals above the ED threshold for the WiFi signals, $\threshToDetectWiFi$\,dBm.
Similarly, a WiFi AP detects the existence of an LTE-U BS in its neighborhood if the AP receives an LTE-U signal above ED threshold $\threshToDetectLTE$\,dBm.
Note that $\threshToDetectWiFi$ and $\threshToDetectLTE$ do not need to be equal.
We denote the bandwidth of an unlicensed channel by $\bandwidth$. 
Transmission power of LTE-U and WiFi is denoted by $P_{\lte}$ and $P_{\wifi}$.
The distance-dependent  
pathloss parameter $\gamma$ is assumed to be identical as both networks are deployed in the same environment and operate at the same frequency.

We assume that LTE BS is equipped with an antenna array of $\numLTEantenna$ antennas~(uniform linear array, ULA) whereas all its users and all WiFi nodes~(i.e., AP as well) have only single antenna.
The LTE BS is able to precode its DL signal for the purpose of beamforming and interference-nulling toward its own UEs as well as a subset of the WiFi nodes to cancel out its interference on these users.
Moreover, we assume the existence of a Cross-Technology Control~(CTC) channel between LTE-U BS and WiFi AP, e.g., LtFi~\cite{gawlowicz2017ltfi}, which is used for exchanging signalling and control data needed for interference nulling. Moreover, it is used for proximity detection, i.e. gives information about the pair of nodes, LTE and WiFi, in mutual interference range.
To compute the precoding matrix for interference-nulling, the LTE-U BS requires knowledge of the channel matrix $H$ towards the WiFi nodes~(refer Section~\ref{sec:nulling}). 
We assume that the LTE-U BS acquires the CSI $H$ information from the control channel. 
We leave how this information can be collected on a practical setting to a future work.

We define the WiFi nodes being nulled by the LTE-U BS as $\usersSetWiFi_{\nullPrefix}$ and their number by $\numLTEantenna_{\nullPrefix}$, i.e., $|\usersSetWiFi_{\nullPrefix}|{=}\numLTEantenna_{\nullPrefix}$.
Denote the LTE BS's beam and nulling configuration $(\userAngle, \usersSetWiFi_{\nullPrefix})$
where $\userAngle$ is the angle between the LTE-U BS and its UE that is being served at this timeslot.
Based on the beamforming/nulling algorithm applied, we can calculate the gain at each user. 
Let us denote the beamforming gain at the receiver under a configuration $(\userAngle, \usersSetWiFi_{\nullPrefix})$
by 
$\AntennaConfig$ and $\AntennaConfig_{i}$ is the gain at UE $\user_i$.
Note that a WiFi station being nulled, e.g., $\userWiFi_i$ , will have a very small $\AntennaConfig_{i}$ value representing the fact that an efficient nulling algorithm results in very weak LTE-U signal at this user.
Under perfect nulling, $\AntennaConfig_{i}$ approaches to zero.

% we define the optimization problem for nulling in LTE+WiFi system

\section{Optimal Interference-Nulling and Beamforming in the LTE-U DL} \label{sec:proposal}

\subsection{Overview}% of \sysname}
As LTE-U does not implement LBT, it has to rely on duty-cycling which is adapted according to the observed WiFi medium utilization and number of WiFi nodes.
Briefly, LTE-U must leave the medium for WiFi proportional to the number of WiFi nodes observed in the neighborhood.
That means, LTE-U's airtime is lower in case of high number of WiFi nodes in the ED range of the LTE-U BS.
Given this key fact as our ground, an LTE-U BS can look for ways to decrease number of WiFi nodes that will be affected by the interference of the LTE-U transmission, i.e., WiFi nodes in its ED range.
This can be achieved in several ways, e.g., decreasing the LTE-U BS transmit power~\cite{chaves2013lte} or handovering some WiFi users equipped with dual radio to the LTE cell~\cite{chen2016rethinking}.
Our approach is different as defined in the following.

We apply cross-technology interference-nulling from LTE-U BS towards carefully-selected WiFi nodes.
As a result, these nulled WiFi nodes receive only very weak interference from the LTE-U DL. 
Hence, from the perspective of the WiFi node, the LTE-U BS is no longer in the competition for the shared medium. 
As a consequence, there is no need to consider such nodes in the estimation of the fair airtime share at the LTE-U BS.
Therefore, LTE-U can maintain a larger share of airtime compared to the case where there is no interference-nulling.
Moreover, since these nulled WiFi nodes are able to receive interference-free traffic during LTE-U's on-period, this approach promises benefits also to the WiFi network. 
On the other hand, longer airtime is achieved at the expense of reserving some of the LTE-U BS's antennas for interference-nulling rather than using it to improve LTE-U's own DL transmission. 
In other words, some of the LTE-U BS's antenna diversity~(aka degree of freedom) is sacrificed for longer airtime usage.
Hence, LTE-U BS needs to apply interference-nulling cautiously, i.e., we need to find the optimal operation point where both networks will be better off.

There are several questions we must address in deriving the optimal operation point: (i) How many of the degrees of freedom, i.e., antennas, an LTE-U BS should use for interference-nulling? (ii) Which of the co-located WiFi nodes~(APs and STAs) should be nulled? To address the above-listed questions which are nontrivial, we need to derive the trade-off between the additional airtime LTE-U gains from interference-nulling and the performance degradation in the LTE-U cell due to the reduced number of degrees of freedom.  
For the first question, we need to formulate the LTE-U throughput considering the airtime as well as the SNR at the UE before and after nulling.
Regarding the second question, the network geometry, i.e. the locations of the co-located WiFi nodes need to be considered, e.g., their distances from the interfering nodes and the serving node~(LTE-U BS or WiFi AP).

Our aim is to find the beamforming/nulling configuration for the LTE-U BS that provides a good balance between the LTE-U and WiFi throughput, which is crucial to achieve a harmonious coexistence in the considered unlicensed bands.
As throughput is a function of the airtime available to a system and the average rate when the considered system captures the medium, 
we explain in the following sections how we calculate the airtime and DL rate of LTE-U and WiFi systems under a particular beamforming/nulling configuration $(\userAngle, \usersSetWiFi_{\nullPrefix})$.
Next, we formulate our problem as a sum-rate maximization problem considering the constraints of the nulling and WiFi-LTE-U coexistence setting.

\subsection{Medium Access under Nulling}\label{sec:medium-access}

Let us explain how nulling may affect the medium access of the WiFi nodes.
Consider a case where all nodes are in a single collision domain.
Since we consider only the DL, WiFi AP and LTE-U BS are the candidate transmitters in this setting who need to apply time sharing.
In case LTE-U BS nulls the WiFi stations~(receivers of WiFi DL traffic), it achieves a higher airtime resulting in lower airtime for the WiFi network.
However, as WiFi AP will defer during LTE-U on-periods, it will not be able to transmit to the nulled WiFi stations in the DL.
In other words, in this case, the WiFi will not benefit from nulling.
However, LTE-U BS can choose to put a null also in the direction of the WiFi AP.
In this case, WiFi AP can transmit all the time and may achieve good channel rate at the nulled stations, if any. 
Nulling only the WiFi stations can improve the WiFi performance in case the WiFi AP is sufficiently far away from the LTE-U BS such that it does not sense the LTE-U BS but WiFi stations are closer to the LTE-U BS. Hence, WiFi DL traffic will benefit from the absence of co-channel interference. Nulling is especially beneficial in a scenario with cross-technology hidden-terminal problem.
In this case, the WiFi AP can send DL traffic to the nulled stations during LTE-U's on-period without LTE interference.

\begin{figure}[!htb]
\includegraphics[scale=0.27]{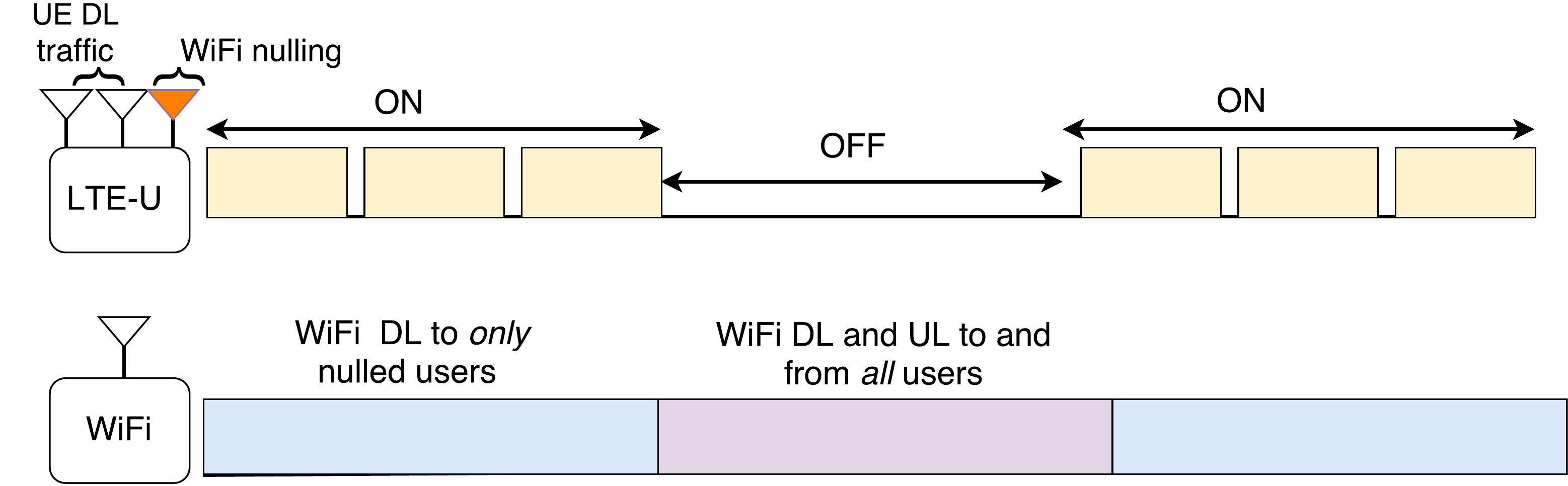}
\caption{Medium access of the LTE-U BS and WiFi nodes.}
\vspace{-10pt}
\label{fig:mediumaccess}
\end{figure}

Fig.~\ref{fig:mediumaccess} shows the medium access in these two considered cases.
While WiFi transmission in both uplink~(UL) and DL could be possible during the LTE-U on-period, it is impossible for LTE-U BS to determine which WiFi node is currently transmitting due to the random access nature of WiFi.
Hence, from a practical point of view, we need a solution where the nulling configuration does not depend on the traffic of the WiFi network but rather only on the positions of the WiFi nodes. We suggest to focus on the WiFi DL which is meaningful as it represents the lion share of the traffic in the WiFi cell.
Therefore, during the LTE-U's on-period, only WiFi DL traffic is considered and any WiFi UL traffic might experience high co-channel interference from LTE-U in case the WiFi AP is not being nulled.\footnote{This will surely create a problem for control frames like immediate ACKs. We recommend to use delayed block acknowledgements available since 802.11n. These frames are sent via contention-based access and therefore can be postponed to the off-period where all types of traffic is possible.}

\subsection{Airtime under Nulling}\label{sec:airtime}

Airtime is the fraction of time a node can access the medium.\footnote{Another term used in the literature for airtime is channel access probability~\cite{cano2015coexistence}.} 
Let us denote by $\airtime_{\lte}$ and $\airtime_{\wifi}$ the LTE-U airtime and WiFi airtime.
Since we consider DL, there is only one transmitter at each network, i.e., LTE-U BS and WiFi AP, we can safely use the term LTE-U airtime or WiFi airtime to refer to the airtime of the LTE-U BS and WiFi AP, respectively.

To calculate airtime at each system, we first need to check if the respective transmitter, BS or AP, senses the other transmitter.
Let $\csRangeFlag_{\wifi}$ represent whether WiFi AP receives the LTE-U BS signal above the predetermined ED level under a beam configuration $\AntennaConfig$.
Please note that WiFi and LTE-U may apply different ED thresholds for signal detection.
We define $\csRangeFlag_{\wifi}$ as follows:
\begin{align}
\csRangeFlag_{\wifi} = 
\begin{cases} 
1 & ,\textrm{ }\frac{P_{\lte}\distanceWiFiLTE^{-\pathlossCoef}\AntennaConfig_{0}}{\bandwidth\eta_0} \geqslant \threshToDetectLTE \label{eq:carrier_sense_flag_WIFI}\\
0 & \textrm{,   otherwise.}
\end{cases}
\end{align}
In (\ref{eq:carrier_sense_flag_WIFI}), we include the term $\AntennaConfig_{0}$ to represent the resulting LTE-U BS's antenna gain at the AP under $\AntennaConfig$, i.e. precoding.

In case $\csRangeFlag_{\wifi}=0$, WiFi's airtime is 1 meaning that it accesses the medium all the time since from its perspective there is no other transmission in the channel requiring it to defer from the channel.
On the other hand, for $\csRangeFlag_{\wifi}=1$, since WiFi applies CSMA-based medium access, the available airtime for WiFi depends on the time the LTE-U does not use the medium, i.e., off-periods.
Hence, we need to first calculate LTE-U's airtime.

LTE-U applies CSAT as the main coexistence scheme. 
Based on the CSAT \textit{on} and \textit{off} periods, we can calculate the airtime for LTE-U simply as $\airtime_{\lte}=\frac{T_{on}}{T_{csat}}$ where $T_{csat} = {T_{on}+T_{off}}$ is the CSAT cycle set to a predefined recommended value, e.g., 80 ms~\cite{QualcommLTEPres15}. 
While there are different suggestions to adapt the CSAT \textit{on} duration~(hence the $T_{off}$ duration as $T_{csat}-T_{on}$),
we will consider the approach suggested in \cite{QualcommLTEPres15} which adapts $T_{on}$ in several iterations according to the medium utilization of WiFi.

Let us now overview the proposal in \cite{QualcommLTEPres15}.
LTE-U small cells are scheduled to sense for WiFi packets during monitoring slots~(in CSAT off period) and estimate the medium utilization~(MU) according to the decoded packet type and its duration.
To ensure a correct estimation of WiFi MU, small cells are all required to stop their transmissions.     %
Given that off-period is sufficiently long, LTE-U cells may perform medium sensing several times and have a better observation about the ongoing WiFi traffic activity.
In our model, we assume backlogged DL for both networks. 
Hence, WiFi's medium utilization converges to 1.

An MU value higher than a threshold, e.g., MU$_1$, triggers LTE-U BS to decrease its $T_{on}$ according to the following equation:
\begin{align}
	T_{on} = \max(T_{on} - \Delta T_{down}, T_{on,min}), \label{eq:Ton}
\end{align}
where $T_{down}$ is the granularity of decrease at each adaptation step and $T_{on,min}$ is the minimum duration for on period to ensure that LTE-U BS can transmit for some minimum duration.
This minimum duration is computed according to the number of WiFi nodes being detected from the preambles of WiFi packets sensed by the LTE-U BS such that the airtime available to each system is \textit{fair}.

\begin{table*}[t]
	\centering
	\renewcommand{\arraystretch}{1.6} 
	\caption{Airtime of LTE-U and WiFi for various CSR($\csRangeFlag_{\wifi},\csRangeFlag_{\lte}$) scenarios: $\csRangeFlag_{x}=1$ means that network $X=\{\lte,\wifi\}$ senses the other network above the ED level. Shaded cell corresponds to the airtime for WiFi when nulling is not applied.}
	\label{tab:airtime}
	\begin{tabular}{|l|l|l|l|l|l|l|}
		\hline
		\multirow{2}{*}{Nw.} &\multirow{2}{*}{CSR(0,0)} & \multirow{2}{*}{CSR(0,1)} & \multirow{2}{*}{CSR(1,0)} &
		\multicolumn{3}{c|}{CSR(1,1)} \\ \cline{5-7}
		&  & & & Null AP   & Null $\numLTEantenna_{\nullPrefix}$ STAs & No Null \\ \hline
		WiFi AP &  1 & 1  &  1-$\frac{1}{\numUsersWiFi_{cs}-\numLTEantenna_{\nullPrefix} + 1}$  & 1     &   1-$\frac{1}{\numUsersWiFi_{cs} - \numLTEantenna_{\nullPrefix}+ 1}$   &  \cellcolor{gray!25} 1-$\frac{1}{\numUsersWiFi_{cs}+ 1}$     \\ \hline
		LTE-U BS  & \multicolumn{6}{c|}{$\frac{1}{\numUsersWiFi_{cs}-\numLTEantenna_{\nullPrefix}+1}$}  \\ \hline
	\end{tabular}
\end{table*}

Let $\numUsersWiFi_{cs}$ denote the number of nodes whose received signal level is above the carrier sense threshold at the LTE-U BS.
We can calculate $\numUsersWiFi_{cs}$ as follows. 
With a slight abuse of the notation, we denote by $\csRangeFlag_{\lte,i}$ the flag taking value 1 if LTE-U BS senses WiFi user $\userWiFi_i$.
	
\begin{align}
\csRangeFlag_{\lte,i} = 
    \begin{cases} 
    1 & ,\textrm{ }\frac{P_{\wifi,i}\distance_{i,\lte}^{-\pathlossCoef}}{\bandwidth\eta_0} \geqslant \threshToDetectWiFi \label{eq:carrier_sense_flag_LTE}\\
    0 & \textrm{,   otherwise}
    \end{cases}
\end{align}
where $P_{\wifi,i}$ is the transmission power of $\userWiFi_i$. 
Consequently, we can compute $\numUsersWiFi_{cs}$ as:
\begin{align}
    \numUsersWiFi_{cs} = \sum_{i=0}^{\numUsersWiFi} \csRangeFlag_{\lte,i}.
\end{align}

\noindent After calculating $\numUsersWiFi_{cs}$, LTE-U can compute $T_{on,min}$ as:
\begin{align}
    T_{on,min} = \min(T_{min}, \frac{(\numUsersLTE_{same}+1)T_{csat}}{\numUsersLTE_{same} + 1 + \numUsersLTE_{other} + \numUsersWiFi_{cs}}), \label{eq:Tonmin}
\end{align}
	where $T_{min}$ is a configuration parameter tuning the minimum duty cycle below ED, $\numUsersLTE_{same}$ is the number of detected LTE-U small cells of the same operator, and $\numUsersLTE_{other}$ is the number of detected small cells of other operators. 
	Note that LTE-U small cells belonging to the same operator have the same public land mobile network ID.
	In the above equation, setting $\numUsersLTE_{same}=0$ and $\numUsersLTE_{other}=0$, we calculate the second term of (\ref{eq:Tonmin}) as 
	$\frac{T_{csat}}{\numUsersWiFi_{cs}+1}$.
	As a smart decision from the perspective of LTE-U is to set $T_{min}$ larger than  $\frac{T_{csat}}{\numUsersWiFi_{cs}+1}$, we can articulate that $T_{on,min}$ is determined by the second term of (\ref{eq:Tonmin}).
Hence, we assume that $T_{on,min} = \frac{T_{csat}}{\numUsersWiFi_{cs}+1}$.

At each iteration of CSAT adaptation, LTE-U BS will be forced to decrease its on duration by $T_{down}$ as in (\ref{eq:Ton}) due to the fact that AP has always DL traffic, i.e., MU $\geqslant$ MU$_1$.
	As a consequence, $T_{on}$ value converges to $T_{on,min}$ which is calculated as $\frac{T_{csat}}{\numUsersWiFi_{cs}+1}$.
	Finally, we can calculate the LTE-U airtime in case of no nulling as:	
	\begin{align}
	\airtime_{\lte}(\numLTEantenna_{\nullPrefix}=0)  = \frac{\frac{T_{csat}}{\numUsersWiFi_{cs}+1}}{T_{csat}} =  \frac{1}{\numUsersWiFi_{cs}+1}.
	\end{align}
	
	If $\numLTEantenna_{\nullPrefix}$ users are nulled, the LTE-U airtime can be calculated as follows:
	\begin{align}
	\airtime_{\lte}(\numLTEantenna_{\nullPrefix}) = \frac{1}{(\numUsersWiFi_{cs} - \numLTEantenna_{\nullPrefix}) + 1}.
	\end{align}
    In the above formula, nulled nodes are neglected while calculating the airtime as they will only marginally be affected by an LTE-U signal under an efficient null steering scheme. Therefore, they become irrelevant in fairness consideration.

Revisiting the case $\csRangeFlag_{\wifi}=1$, we can calculate WiFi airtime depending on whether LTE-U BS nulls the AP or not.
Note that nulling the AP is no different than nulling a WiFi station and can be considered as an option.
In case WiFi AP is nulled, the WiFi airtime equals to 1. 
That is, interference nulling at the WiFi AP results in WiFi AP never defer as it will never sense an ongoing LTE-U transmission.
If LTE-U does not prefer to null the AP, WiFi airtime is simply $\airtime_{\wifi} = 1 - \airtime_{\lte}(\numLTEantenna_{\nullPrefix})$.

\begin{figure}[!htb]
\subfloat[LTE-U airtime for various $\numUsersWiFi_{cs}$.]{\includegraphics[width=.34\textwidth]{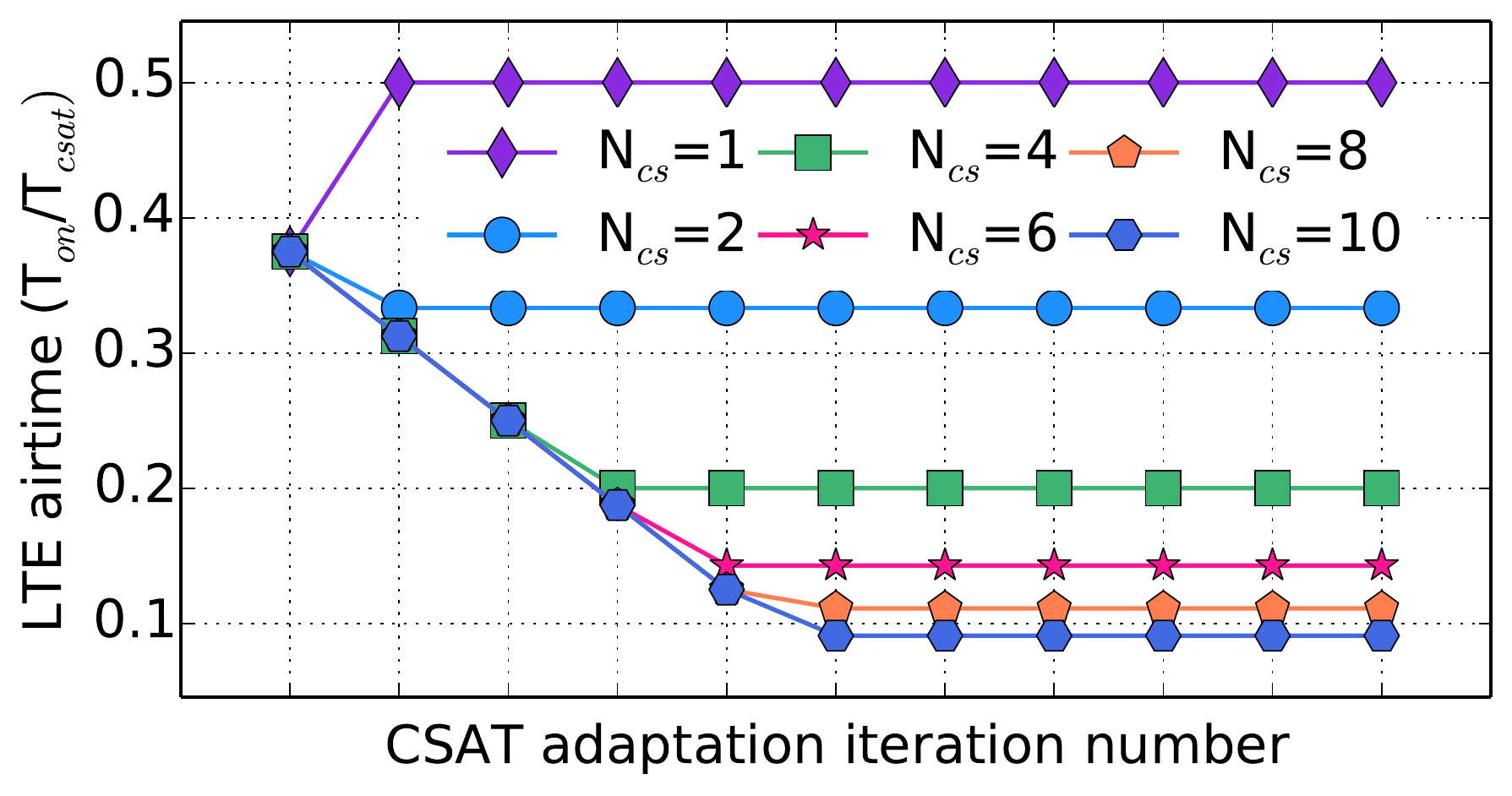}\label{fig:airtime_all_LTE_CSAT}} \hfill
\subfloat[Nulling gain in terms of airtime increase with increasing $\numUsersWiFi_{cs}$ and for various $\numLTEantenna_{\nullPrefix}$.]{\includegraphics[width=.34\textwidth]{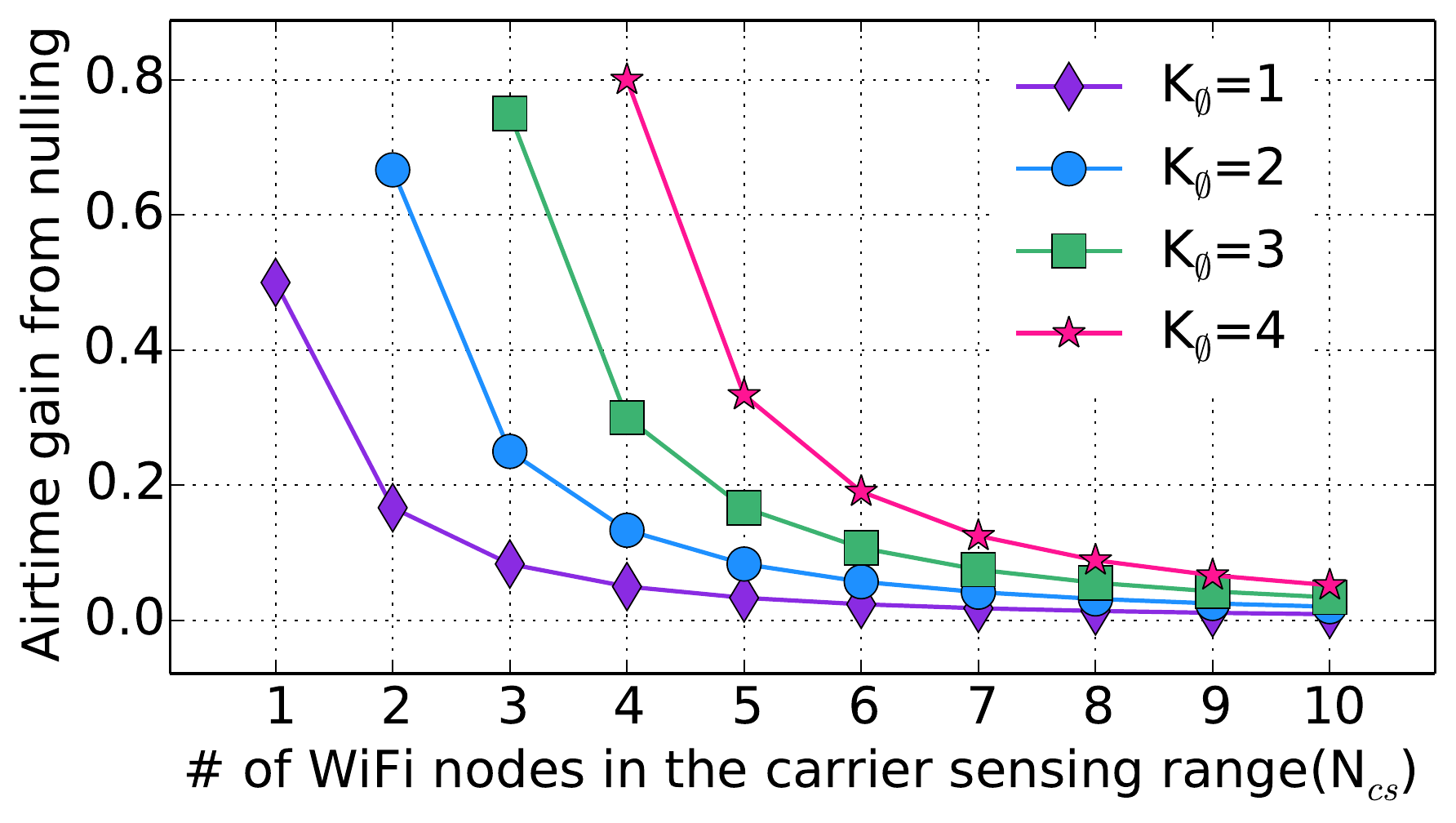}\label{fig:nulling_gain}} 
\caption{Impact of CSAT adaptation and nulling on LTE-U airtime.}
\vspace{-10pt}
\label{fig:analysis_of_csat_nulling} 
\end{figure}

Table~\ref{tab:airtime} summarizes airtime values depending on the carrier sensing condition of each network.
Let CSR($\csRangeFlag_{\wifi}$,$\csRangeFlag_{\lte}$) denote all cases where $\csRangeFlag_{\lte}=\{0,1\}$ and $\csRangeFlag_{\wifi}=\{0,1\}$.
Note that we have the same airtime formula for the LTE-U independent of its $\csRangeFlag_{\lte}$ value as there may be cases where LTE-U does not hear the transmission of the AP but overhears those of the stations.
That is to say, what really matters for LTE-U is the number of WiFi nodes in the ED range of the LTE-U BS. 
Regarding WiFi, we must consider 
$\csRangeFlag_{\wifi}$ as well as the nulling status of the AP.

Implementing the approach of \cite{QualcommLTEPres15}, we find the change in LTE-U airtime at each CSAT adaptation step with increasing $\numUsersWiFi_{cs}$ under the assumption that medium utilization is 1, i.e., WiFi traffic is backlogged.
We set the initial values of $T_{on}=40$ ms, $T_{off}=40$ ms, $T_{csat}=80$ ms, $\Delta T_{down}=5 ms$.
Moreover, we set $T_{on,min}=80$ ms to let LTE-U be constrained by the WiFi traffic not artificially by its misconfiguration.

Fig.\ref{fig:airtime_all_LTE_CSAT} plots the LTE-U airtime, i.e., $\airtime_{\lte}=\frac{T_{on}}{T_{csat}}$, for various number of neighboring WiFi nodes.
Notice that the airtime values converge to $\frac{1}{\numUsersWiFi_{cs}+1}$ after some adaptation steps as expected from our analysis.
The convergence speed obviously depends on the initial value of $T_{on}$ as well as $T_{csat}$, number of WiFi stations in the coexistence domain~($\numUsersWiFi_{cs}$) and how successfully LTE-U can detect their existence~($MU$ and $\numUsersWiFi_{cs}$), and the granularity of decrease/increase steps~($\Delta T_{down}, \Delta T_{up}$).
However, these are beyond the scope of the current paper.
From Fig.\ref{fig:airtime_all_LTE_CSAT}, we can also observe the nulling gain as the difference between the curves corresponding two different $\numUsersWiFi_{cs}$ curves. 
For example, for the initial setting of $\numUsersWiFi_{cs}=10$, we will get the nulling gain in terms of airtime under $\numLTEantenna_{\nullPrefix}=2$ as much as the difference of airtimes for $\numUsersWiFi_{cs}=8$ and that of $\numUsersWiFi_{cs}=10$, i.e., 1/9-1/11.
In case of lower $\numUsersWiFi_{cs}$, the benefit of nulling is more pronounced as shown in Fig.\ref{fig:nulling_gain} which plots the gain in the LTE-U airtime by nulling $\numLTEantenna_{\nullPrefix}$ WiFi nodes under different $\numUsersWiFi_{cs}$ values.

\subsection{Throughput under nulling} % a beamforming and nulling configuration
Let us consider an LTE-U UE and calculate its throughput in the DL.
Recall that the LTE-U BS applies beamforming and the resulting gain at $\user_j$ is denoted by $\AntennaConfig_{j}$ whereas WiFi AP does not as we assume a single antenna at the WiFi AP.
For the LTE-U UE $\userLTE_j$, DL rate can be defined as:
\begin{align}
\rate_{j, \lte} = 
\begin{cases} 
\rate_{j,\lte}^{0} = 
\bandwidth\log(1+\frac{P_{\lte}\distance^{-\pathlossCoef}_{j,\lte}\AntennaConfig_{j}}{\bandwidth\eta_0}),  & \textrm{ blocked WiFi AP}\\
\rate_{j, \lte}^{1} = \bandwidth\log(1+\frac{P_{\lte}\distance^{-\pathlossCoef}_{j,\lte}\AntennaConfig_{j}}{\bandwidth\eta_0 + P_{\wifi}\distance_{j,\wifi}^{-\pathlossCoef}}),& \textrm{ unblocked WiFi AP}
\end{cases}
\end{align}
where WiFi AP may be unblocked in two cases: (i) the AP does not sense LTE-U BS, i.e., $\csRangeFlag_{\wifi}=0$, or (ii) despite $\csRangeFlag_{\wifi}=1$, the AP can transmit because it is nulled. 
Note that in the above equation $\AntennaConfig_{j}$ is a function of the number of antennas used for nulling.
The LTE-U BS uses its ($\numLTEantenna-\numLTEantenna_{\nullPrefix}$) antennas for this UE resulting in lower beam gain if less antennas are available for the UE.
As we already calculated the airtime for LTE, we can find the throughput for an LTE UE as: $\throughput_{j,\lte} = \airtime_{\lte}\rate_{j,\lte}$.

As for WiFi DL rate, we need to consider whether coexistence is only in the time domain or in both time and space domains. 
For the former, there will be no LTE-U BS interference on the WiFi DL.
However, for the latter, as LTE-U BS changes state between on and off periods while WiFi AP has DL traffic, 
we calculate the WiFi DL rate at WiFi station $\userWiFi_i$ considering the rates during on and off periods.
Let us consider the first case, i.e., l{$\csRangeFlag_{\wifi}=1$ and AP is not nulled. WiFi throughput in this case $\throughput_{i, \wifi}^{0}$ equals to:
\begin{align}
	\throughput_{i, \wifi}^{0}= (1-\airtime_{\lte})\bandwidth\log(1+\frac{P_{\wifi}\distance^{-\pathlossCoef}_{i,\wifi}}{\bandwidth\eta_0}).
	\end{align}
If sharing is in time and space, i.e., $\csRangeFlag_{\wifi}=0$ or AP is nulled, WiFi throughput $\throughput_{i, \wifi}^{1}$ equals to:
\begin{align} 
\throughput_{i, \wifi}^{1}=
	 \underbrace{\airtime_{\lte}\bandwidth\log(1+\frac{P_{\wifi}\distance^{-\pathlossCoef}_{i,\wifi}}{\bandwidth\eta_0 + P_{\lte}\distance_{i,\lte}^{-\pathlossCoef}\AntennaConfig_{i}})}_{\text{LTE on-period}} + \underbrace{(1-\airtime_{\lte})\bandwidth\log(1+\frac{P_{\wifi}\distance^{-\pathlossCoef}_{i,\wifi}}{\bandwidth\eta_0})}_{\text{LTE off-period}}.
\end{align}
Note that if $\userWiFi_i$ is in $\usersSetWiFi_{\nullPrefix}$, $\AntennaConfig_{_i}$ is marginal and effectively results in no rate degradation in the WiFi DL for $\userWiFi_i$.

\subsection{Channel access delay under nulling}\label{sec:ch_access_delay}
Let us now calculate the expected time to access the medium for both LTE-U BS and the WiFi AP considering the case of CSR($\csRangeFlag_{\wifi}=1$,\,$\csRangeFlag_{\lte}=1$).
In a conventional LTE-U/WiFi setting, the LTE-U BS has to wait for the on-period to be able to send its packets while WiFi AP waits for the LTE-U off-period. 
In this case, expected channel access delay for LTE-U BS $\delay_{\lte}$ is:
\begin{align}
\delay_{\lte} = (1{-}\airtime_{\lte})\frac{T_{off}}{2}{=} (1-\airtime_{\lte})\frac{(1-\airtime_{\lte})T_{csat}}{2}{=}\frac{(1{-}\airtime_{\lte})^2T_{csat}}{2} \label{eq:airtime_lte}
\end{align}
Similarly, we calculate the expected channel access delay for WiFi AP $\delay_{\wifi}$ as\footnote{Note that we neglect the subframe punctures in this calculation considering LTE-U on periods shorter than 20 ms.}:
\begin{align}
\delay_{\wifi} = \airtime_{\lte}\frac{T_{on}}{2}= \airtime_{\lte}\frac{\airtime_{\lte}T_{csat}}{2}=\frac{\airtime_{\lte}^2T_{csat}}{2}. \label{eq:airtime_wifi}
\end{align}
Under nulling, LTE-U BS experiences a faster access to the channel as LTE airtime $\airtime_{\lte}$ is increased.
For WiFi, channel access delay gets shorter if AP is nulled: 
essentially we move from the regime of CSR(1,1) to that of CSR(0,1).
As a result, channel access delay becomes zero for the WiFi AP.

\subsection{Problem Formulation}\label{sec:problem}
Our aim is to find the nulling configuration to be used at the LTE-U BS that provides the \textit{optimal} performance.
We can define different optimization objectives by changing the priority of LTE-U and WiFi denoted by $\priority_{\lte}$ and $\priority_{\wifi}$ and satisfying the condition that $\priority_{\lte}+\priority_{\wifi}=1$. Our policies are:
\begin{itemize}
	\item \textrm{MaxSum} aims at maximizing the system wide capacity giving each system equal weight~(i.e., $\priority_{\lte}=\priority_{\wifi}=0.5$) with a constraint that WiFi capacity does not degrade compared to the baseline in which LTE-U does not apply nulling~(referred to as \nonull).
	\item \textrm{MaxLTE} aims at maximizing LTE-U's capacity, i.e., $\priority_{\lte}=1$ and $\priority_{\wifi}=0$.
	\item \textrm{MaxWiFi} aims at maximizing WiFi's capacity, i.e., $\priority_{\wifi}=1$ and $\priority_{\lte}=0$.
\end{itemize}
Let $\mathbf{x}=[x_0, \cdots, x_N]$ denote the LTE-U BS's nulling configuration where
$x_i$ yields value 1 if WiFi station $i$ is nulled and 0 otherwise.  We can formulate our optimization problem as follows:
\begin{subequations}
\begin{eqnarray}
&\max  \quad \priority_{\wifi}\frac{ \sum_{i=1}^{\numUsersWiFi}\throughput_{i,\wifi}}{\numUsersWiFi}+
\priority_{\lte} \airtime_{\lte} \sum_{j=1}^{\numUsersLTE} \rate_{j,\lte} \label{obj} \\
&\textrm{subject to} \nonumber \\
&\throughput_{i,\wifi}{=}\csRangeFlag_{\wifi}((1{-}x_o)\throughput_{i,\wifi}^{0}{+}x_0\throughput_{i,\wifi}^{1}) {+} (1{-}\csRangeFlag_{\wifi})\throughput_{i,\wifi}^{1},\, \forall i{=}[1,\numUsersWiFi] \label{const:wifi_rate}\\
&\rate_{j, \lte}{=}y_j(x_0\rate_{j,\lte}^{1} {+} (1{-}x_0)(\csRangeFlag_{\wifi}\rate_{j,\lte}^{0} {+} (1-\csRangeFlag_{\wifi})\rate_{j,\lte}^{1})),\, \forall j\,=\,[1,\numUsersLTE] \label{const:lte_rate}  \\
&\sum_{j=1}^{\numUsersLTE} y_j=1 \label{const:one_UE_active}\\
&x_i \leqslant \csRangeFlag_{\lte, i}, \quad \forall i=[0,\numUsersWiFi] \label{const:null_only_nodes_in_csrange}\\
&\sum_{i=1}^{\numUsersWiFi} x_i<\numLTEantenna \label{const:num_antenna_constraint}\\
&\airtime_{\lte} = \frac{1}{\sum_{i=0}^{\numUsersWiFi} \csRangeFlag_{\lte,i} - \sum_{i=0}^{\numUsersWiFi}x_i + 1} \label{const:lte_airtime}\\
&\airtime_{\wifi} = x_0 {+} (1{-}x_0)(\csRangeFlag_{\wifi}(1-\airtime_{\lte}) + (1-\csRangeFlag_{\wifi})) \label{const:wifi_airtime}\\
&x_i \in \{0,1\} \quad\quad \forall i=[1,\numUsersWiFi] \label{const:binary_variables} \\
&y_j \in \{0,1\} \quad\quad \forall j=[1,\numUsersLTE] \label{const:y_binary_variables} 
\end{eqnarray}
\end{subequations}

In the above formulation, first term of our objective~(\ref{obj}) represents the expected DL throughput of the WiFi network weighted by $\priority_{\wifi}$ and the second term stands for the throughput of the LTE network weighted by $\priority_{\lte}$.
Consts.~\ref{const:wifi_rate} and \ref{const:lte_rate} correspond to the throughput of a WiFi user and rate of an LTE-U user, respectively.
Binary variable $y_j$ in Const.~\ref{const:lte_rate} represents whether UE $j$ is scheduled to receive DL traffic.
Const~\ref{const:one_UE_active} states the fact that there is only one UE actively receiving DL traffic from the LTE-U BS at any scheduling period.
Since airtime increase is only relevant for nodes that are in the ED range of the LTE-U BS, we add  Constr.~\ref{const:null_only_nodes_in_csrange} to ensure that $x_i$ is zero if $\userWiFi_i$ is not in the range of LTE-U BS.
Such WiFi nodes are not selected for nulling due to Const.\ref{const:null_only_nodes_in_csrange}.
Const.~\ref{const:num_antenna_constraint} states that maximum number of nulled WiFi nodes must be smaller than the total number of LTE-U antennas such that at least one antenna is reserved for its UE.
Consts.\ref{const:lte_airtime} and \ref{const:wifi_airtime} define the airtimes of LTE-U and WiFi, respectively.
Note that $x_0$ in Const.~\ref{const:wifi_airtime} stands for WiFi AP and states the fact that if WiFi AP is nulled, the airtime for WiFi will be 1.
Finally, Consts.\ref{const:binary_variables} and \ref{const:y_binary_variables} denote the type of variables as binary integers.

Note that this problem can be solved for both $\mathbf{x}=[x_i]$ and $\mathbf{y}=[y_j]$ simultaneously or setting $\mathbf{y}$ first it can be solved for $x$.
In this work, we take $\mathbf{y}$ as given, i.e., LTE BS scheduler first decides on which UE to serve.
Nevertheless, the problem of determining $\mathbf{x}$  is of high complexity.
Therefore, in the following we present a low-complexity algorithm which can be implemented easily and run at every duty-cycle period of the LTE-U BS.

%% proposed heuristic
%!TEX root = main.tex
\vspace{-0.2cm}
\section{Low-Complexity Nulling: \heuristicname}\label{sec:greedy}
Randomly selecting the WiFi nodes~(AP or STAs) to be nulled by the LTE-U BS is suboptimal as it may either degrade the wanted signal towards the UE, i.e., in case the WiFi node to be nulled and the LTE-U UE cannot be separated in angular domain~(or two channels are correlated), or nulling a WiFi transmitter may result in hidden terminal problem as the CS mechanism at the WiFi node is effectively switched off due to nulling.
To avoid such cases, we propose a null grouping algorithm that groups WiFi nodes into suitable subsets that are beneficial to null. %
The task of a null grouping algorithm is thus to determine an efficient nulling group with a low complexity.

Our proposed heuristic for estimating the nulling group is a greedy algorithm~(\heuristicname) that constructs a null group starting with the WiFi node that when being nulled gives the largest gain in terms of the selected metric, i.e., largest increase in LTE-U capacity, and sequentially extending this group by admitting the WiFi node providing the highest increase of a given grouping metric~(refer to three policies in Section~\ref{sec:problem}). 
Once the group reaches its target size, or no more WiFi nodes can increase the grouping metric, the nulling group is considered complete.
Note that the following information is needed to compute the metric: i) the set of WiFi nodes~(STAs/AP) in the sensing range of the LTE-U BS, ii) the average pathloss of the channel from WiFi AP towards LTE-UE currently being served.

\smallskip

\noindent \textit{Complexity:} The computational complexity of our heuristic regarding its execution time in terms of number grouping metric calculations is $\mathcal{O}((N+1)^2)$ where $N+1$ corresponds to the number of WiFi nodes\textemdash AP and STAs.

%% Evaluation by network simulations
\section{Performance Evaluation}\label{sec:eval}
\begin{figure*}[htb]
\minipage{0.32\textwidth}
\subfloat[LTE-U throughput.]{\includegraphics[width=\linewidth]{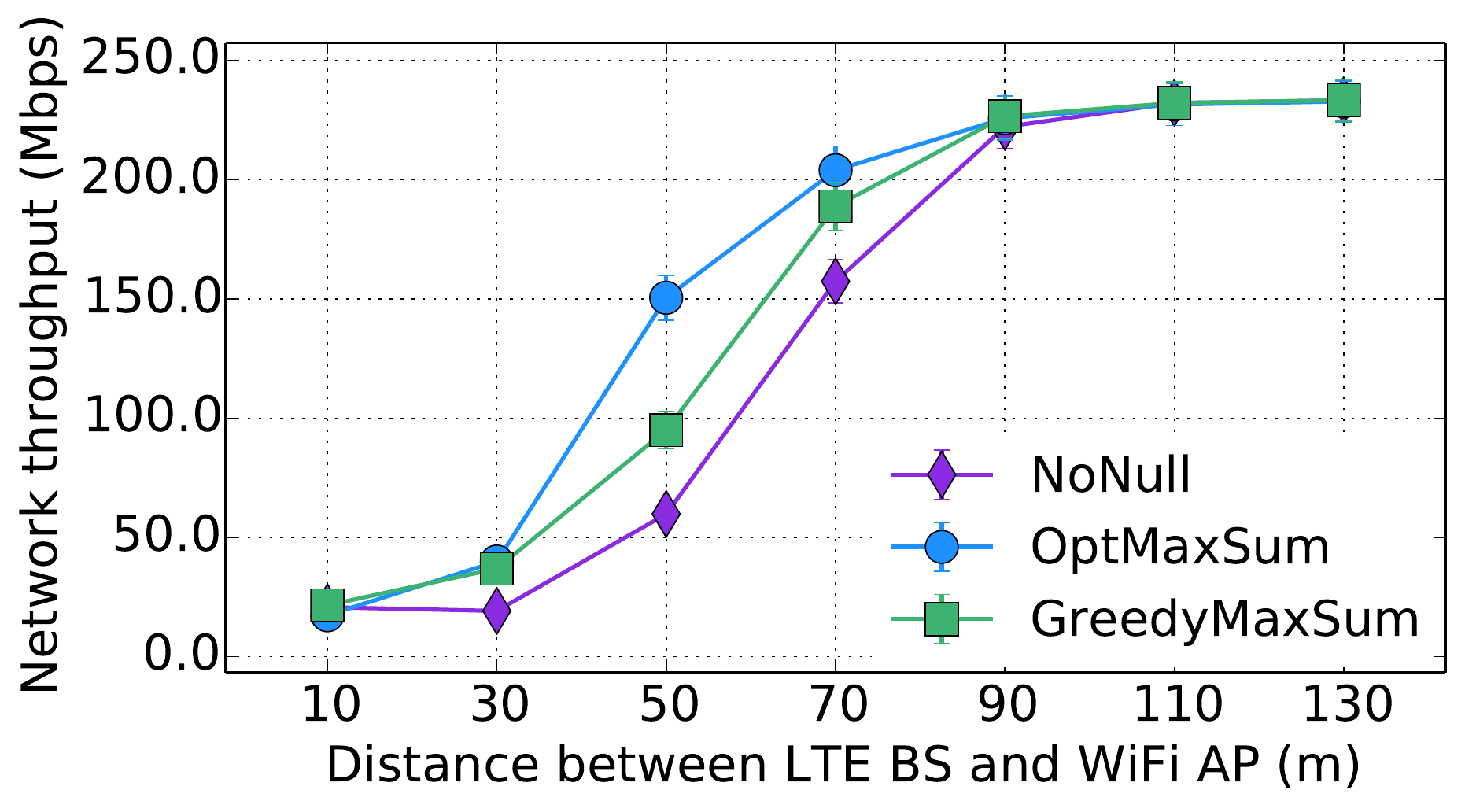}\label{fig:eval_opt_greedy_LTE}} \hfill
\subfloat[WiFi throughput.]{\includegraphics[width=\linewidth]{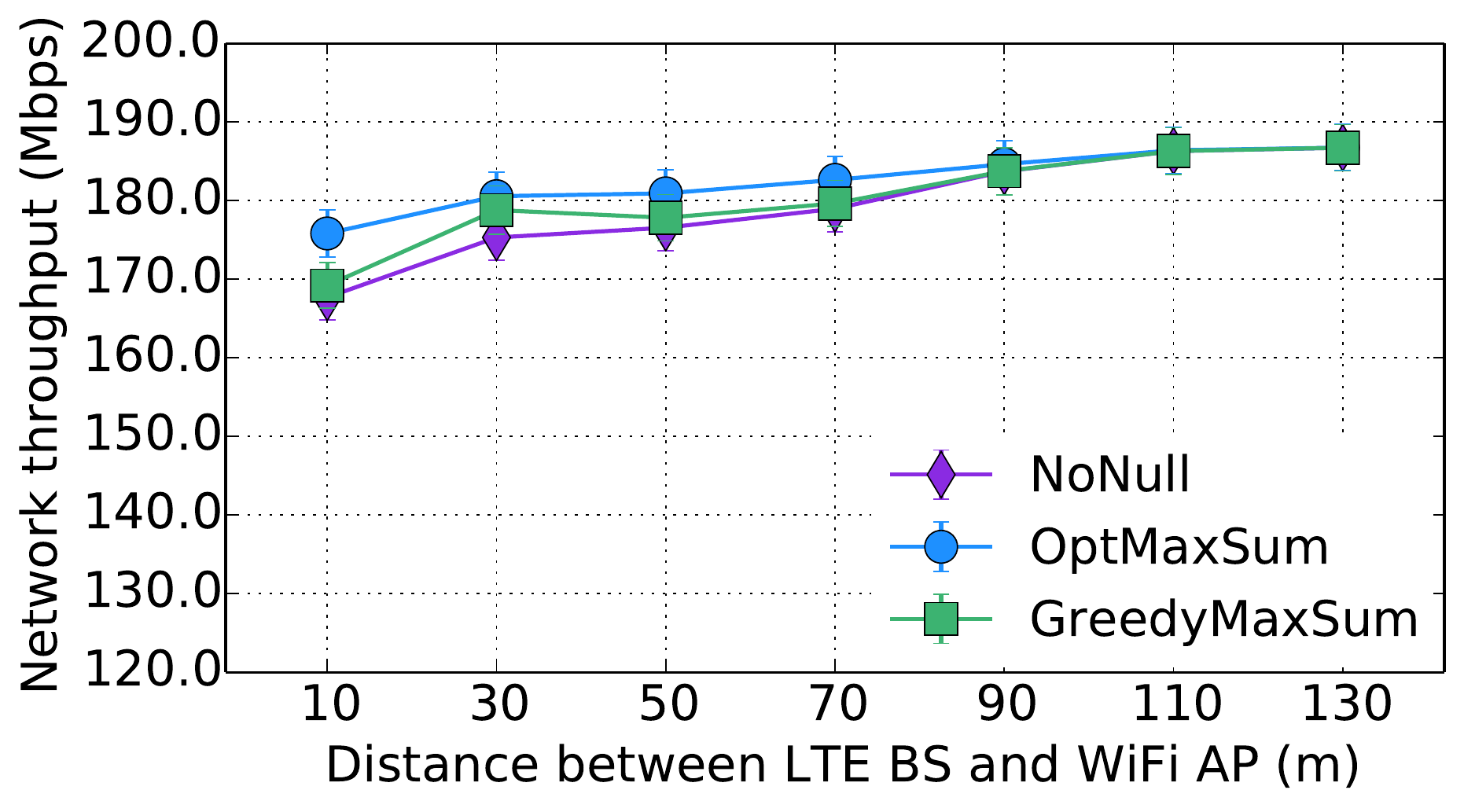}\label{fig:eval_opt_greedy_WiFi}} 
\caption{Comparison of schemes for $\numLTEantenna$=6, $\numUsersWiFi=8$.\label{fig:eval_opt_greedy}}
\endminipage\hfill
\minipage{0.32\textwidth}
  	\subfloat[LTE-U throughput.]{\includegraphics[width=\linewidth]{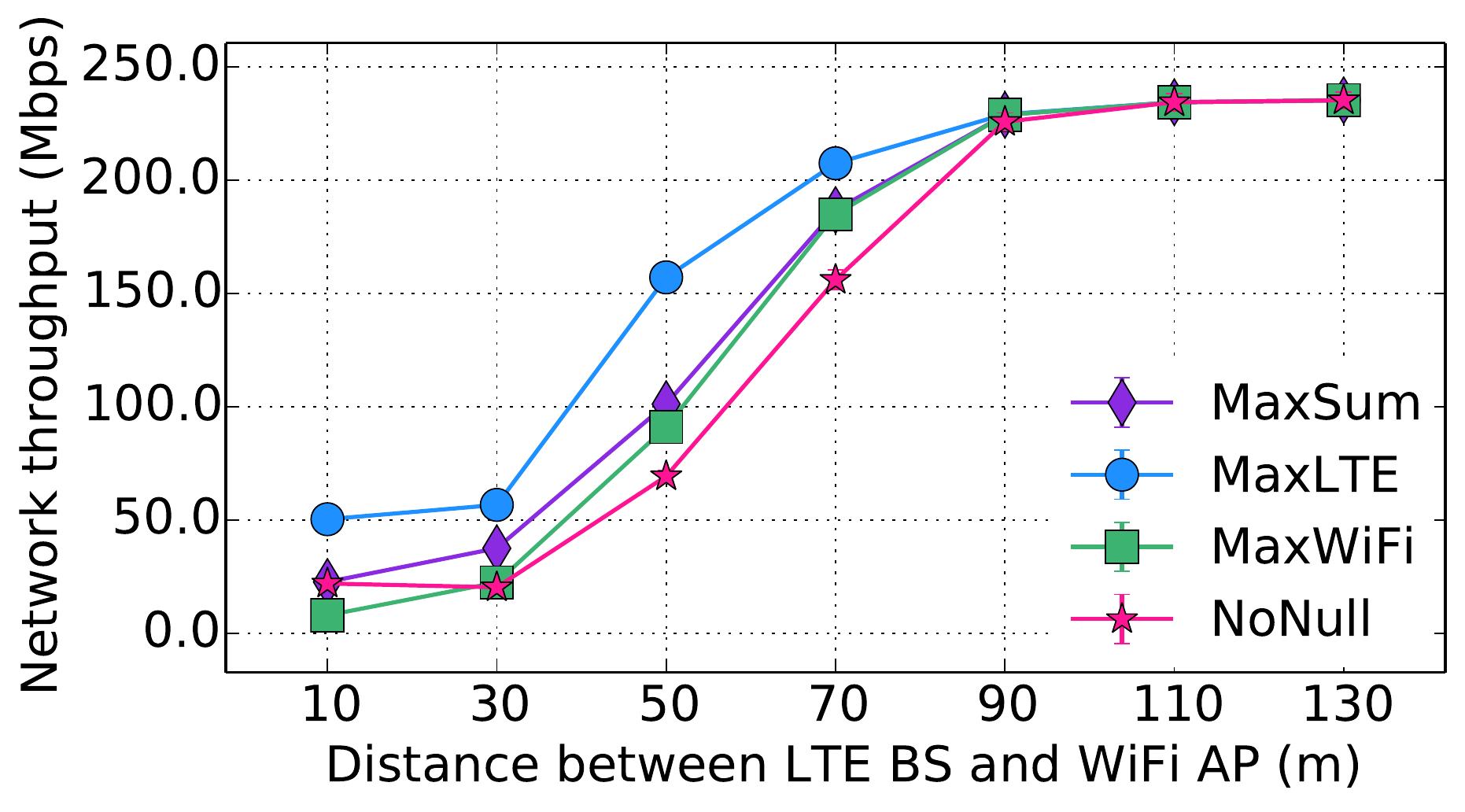}\label{fig:eval_strategies_LTE}} \hfill
  	\subfloat[WiFi throughput.]{\includegraphics[width=\linewidth]{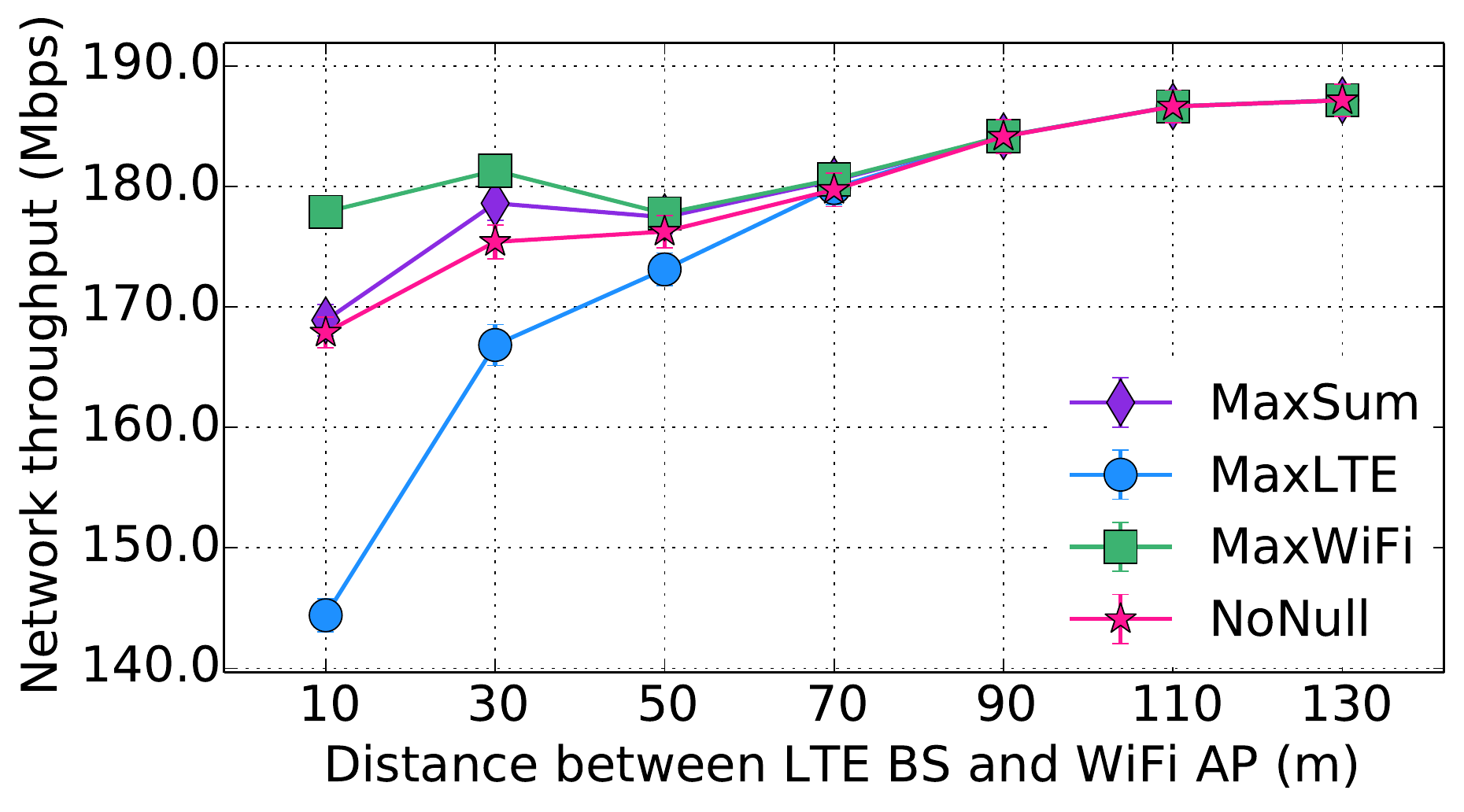}\label{fig:eval_strategies_WiFi}} 
	\caption{Comparison of optimization objectives, $\numLTEantenna$=6, $\numUsersWiFi=8$.\label{fig:eval_strategies}}
\endminipage\hfill
\minipage{0.32\textwidth}%
	\subfloat[LTE-U channel access delay.]{\includegraphics[width=\linewidth]{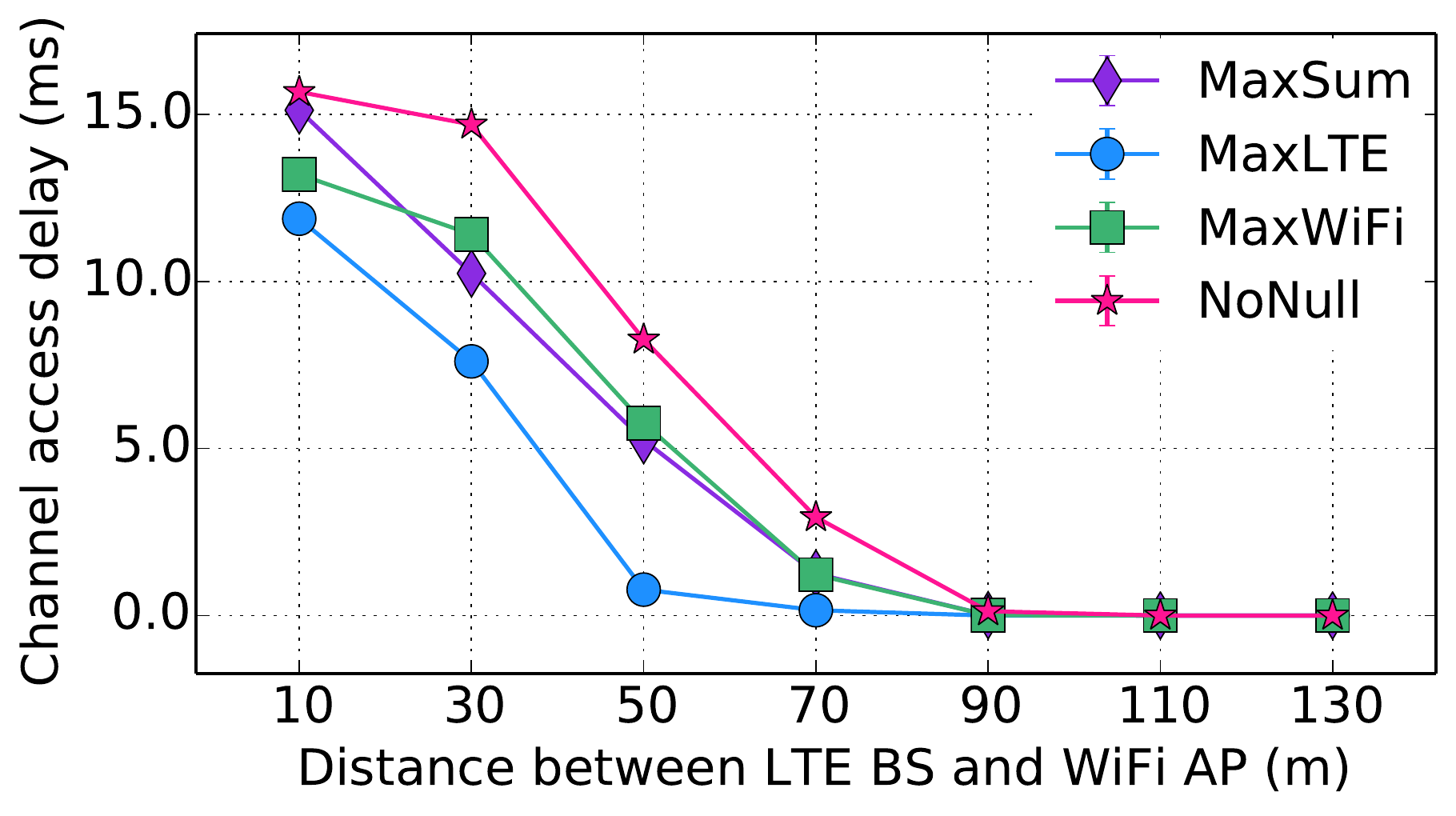}\label{fig:eval_strategies_chacccess_LTE}} \hfill
		\subfloat[WiFi channel access delay.]{\includegraphics[width=\linewidth]{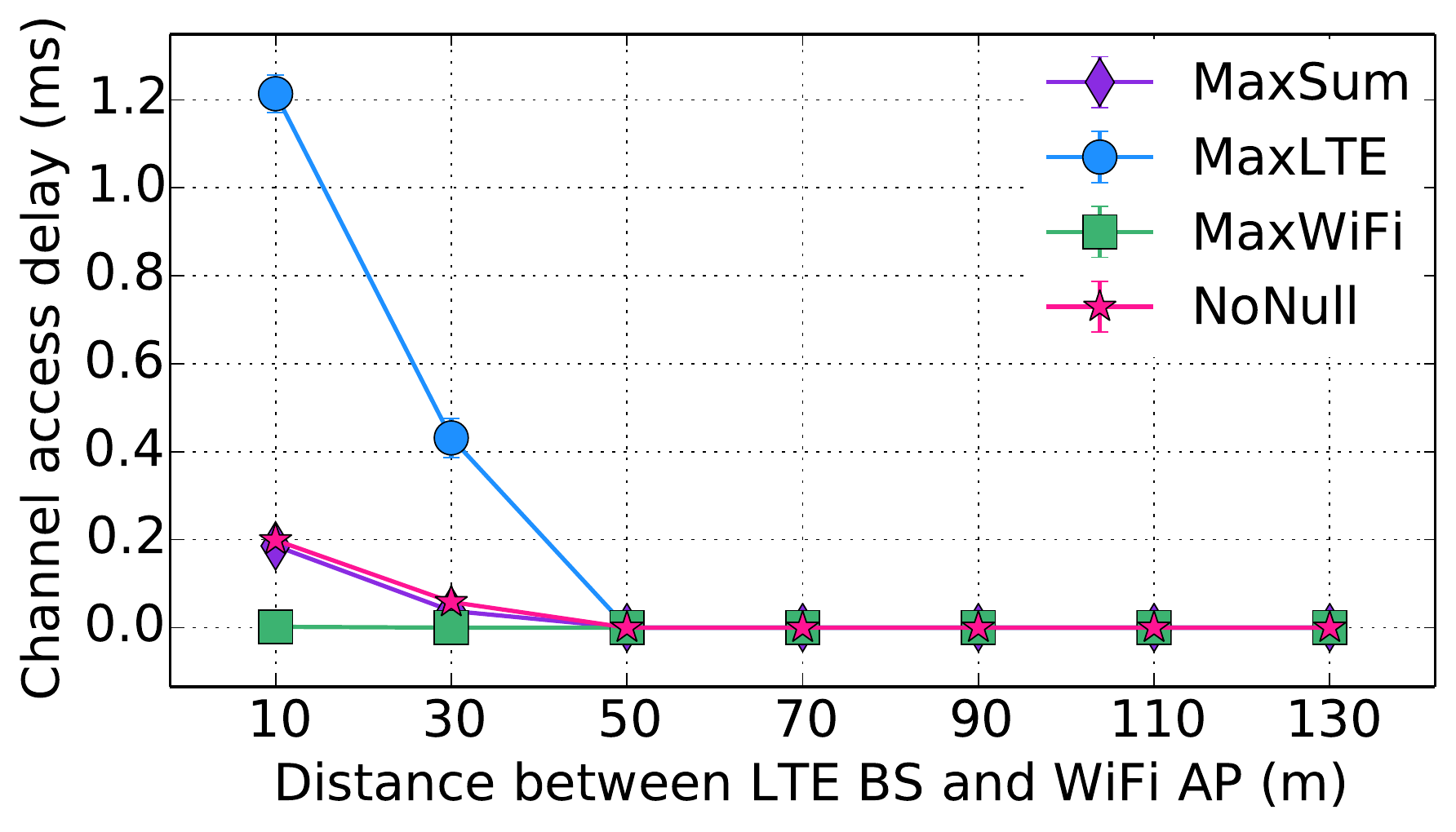}\label{fig:eval_strategies_chacccess_WiFi}} 
		\caption{Channel access delays , $\numLTEantenna$=6, $\numUsersWiFi=8$.\label{fig:eval_strategies_ch_access}}
\endminipage
\end{figure*}

We evaluate our approach by means of network simulations in Python while computing the antenna array response after precoding~(beamforming/nulling) in Matlab's Phased Array system toolbox.\footnote{\url{https://de.mathworks.com/products/phased-array.html}}
Specifically, we derive the precoding vector using LCMV beamformer~\cite{van2002optimum} as it allows us to put the signal in the desired direction~(i.e., UE) while putting nulls towards selected WiFi nodes.

Unless otherwise stated, we use the following parameters: number of UEs $\numUsersLTE$=1, $P_{\lte}$=17\,dBm,  $P_{\wifi}$=17\,dBm as well as the power of WiFi stations while calculating $\numUsersWiFi_{cs}$, $\Gamma_{\wifi}=-82$\,dBm, $\Gamma_{\lte}=-72$\,dBm.  
To determine the location of each user, we randomly select an angle in interval [0, 2$\pi$] and distance in the interval [0, coverage radius] where coverage radius is 50\,m for both LTE and WiFi.
We change $\distanceWiFiLTE$ from 10\,m to 130\,m with a step of 20\,m to cover all interference regimes. 
In the following, we present the mean statistics collected from 500 runs.
The plots also show the standard error of the mean values, which are mostly very small.

\subsection{Gain from Nulling}
In this section, we show how much gain both LTE-U and WiFi network achieves through nulling with our \heuristicname~ algorithm and \textrm{MaxSum} policy.
To evaluate how suboptimal \heuristicname~ is, we also provide the results for optimal solution~(OptMaxSum) found through exhaustive search of all possible nulling groups considering the objective function in (\ref{obj}). 
OptMaxSum maximizes the sum of WiFi and LTE-U throughput.

Fig.~\ref{fig:eval_opt_greedy} compares \nonull~ with the proposed nulling scheme for different distances between the LTE-U BS and WiFi AP for $\numLTEantenna$=6 and $\numUsersWiFi$=8. 
Note that the number of antennas is very moderate and already available in most today's commodity BSs.
As Fig.~\ref{fig:eval_opt_greedy_LTE} depicts, the performance of the LTE-U cell is higher under nulling compared to \nonull.
The increase in throughput is mostly due to the increased LTE-U duty cycle because of nulling. 
The performance increase achieved by \optname~ is up to 152\% for LTE which is realized at $\distanceWiFiLTE$=50\,m.
\heuristicname~ achieves up to 92\% improvement over \nonull~ and the highest gain is realized at $\distanceWiFiLTE$=30\,m.
Second observation worth noticing is that the difference between \heuristicname~ and \optname~ is mostly low with the exception at $\distanceWiFiLTE$=50\,m. 

As of WiFi performance, we observe in Fig.~\ref{fig:eval_opt_greedy_WiFi} that WiFi cell slightly benefits from nulling.
At $\distanceWiFiLTE$=10\,m, the WiFi throughput is increased by 5\%~(and 1\% by \heuristicname) which corresponds to the highest gain for WiFi. 
However, for sparse user deployments, achieved throughput gain is higher. 
For example, for a WiFi cell with a single station~(not plotted), OptMaxSum provides 44\% increase to the WiFi cell at $\distanceWiFiLTE$=10\,m and 19\% increase at $\distanceWiFiLTE$=30\,m.
Corresponding gain for \heuristicname~ is 10\% and 13\%. 
For high distance, e.g., $\distanceWiFiLTE$>90\,m, there is no need for nulling as both networks are already separated in space and their mutual interference approaches to zero.

\vspace{-0.3cm}
\subsection{Impact of optimization objective}

Fig.~\ref{fig:eval_strategies} shows the change in each network's throughput achieved by \heuristicname~ under each nulling policy, i.e., maximize the LTE-U throughput, WiFi throughput, or sum of both networks' throughput. 
We see that \textrm{MaxSum} offers a very good balance between LTE-U and WiFi performances: it achieves nonnegative gains at each network while other two objective might result in one network to suffer. 
Fig.~\ref{fig:eval_strategies_ch_access} shows a similar trend considering the channel access delay of each network for LTE-U $T_{csat}$=40\,ms.
In Fig.\ref{fig:eval_strategies_chacccess_LTE}, we also observe the reduction in the channel access latency at the LTE-U BS facilitated by nulling.
For WiFi AP, channel access is faster than that of LTE-U BS due to longer airtime of the WiFi cell for this setting with $\numUsersWiFi$=8.
Nevertheless, \textrm{MaxWiFi} can decrease it even further toward zero. 
However, considering the LTE-U network's performance, we pick \textrm{MaxSum} as our policy for \heuristicname~ in the following analysis.

\subsection{Impact of number of LTE-U BS antennas}

Fig.~\ref{fig:eval_impact_antenna} shows the impact of the number of LTE BS antennas when the neighboring WiFi cell has 8 stations.
Here, we present the absolute throughput gain of the proposed scheme over \nonull.
Unsurprisingly, we observe in Fig.~\ref{fig:eval_impact_antenna_LTE} that the LTE-U throughput can be increased significantly with larger number of antennas. 
This improvement is due to both increased beamforming gain and the possibility to steer multiple nulls.
With increasing $\distanceWiFiLTE$, we first observe an increasing throughput gain.
In this region, the increase in airtime due to more nulls outweighs the sacrificed antenna diversity at the LTE-U cell. 
As we observed also in Fig.~\ref{fig:eval_impact_antenna_LTE}, with further increase in distance, the need for interference nulling diminishes resulting in no throughput gain.
For example, for $\numLTEantenna$=10, achieved gains are (26\%,\,221\%,\,61\%,\,20\%,\,1\%) for $\distanceWiFiLTE$=(10,\,30,\,50,\,70,\,90)\,m.
Note that having 10 or more antennas is in line with 5th generation mobile networks (5G).

\begin{figure*}[!htb]
\minipage{0.32\textwidth}
  	\subfloat[LTE-U throughput gain.]{\includegraphics[width=\linewidth]{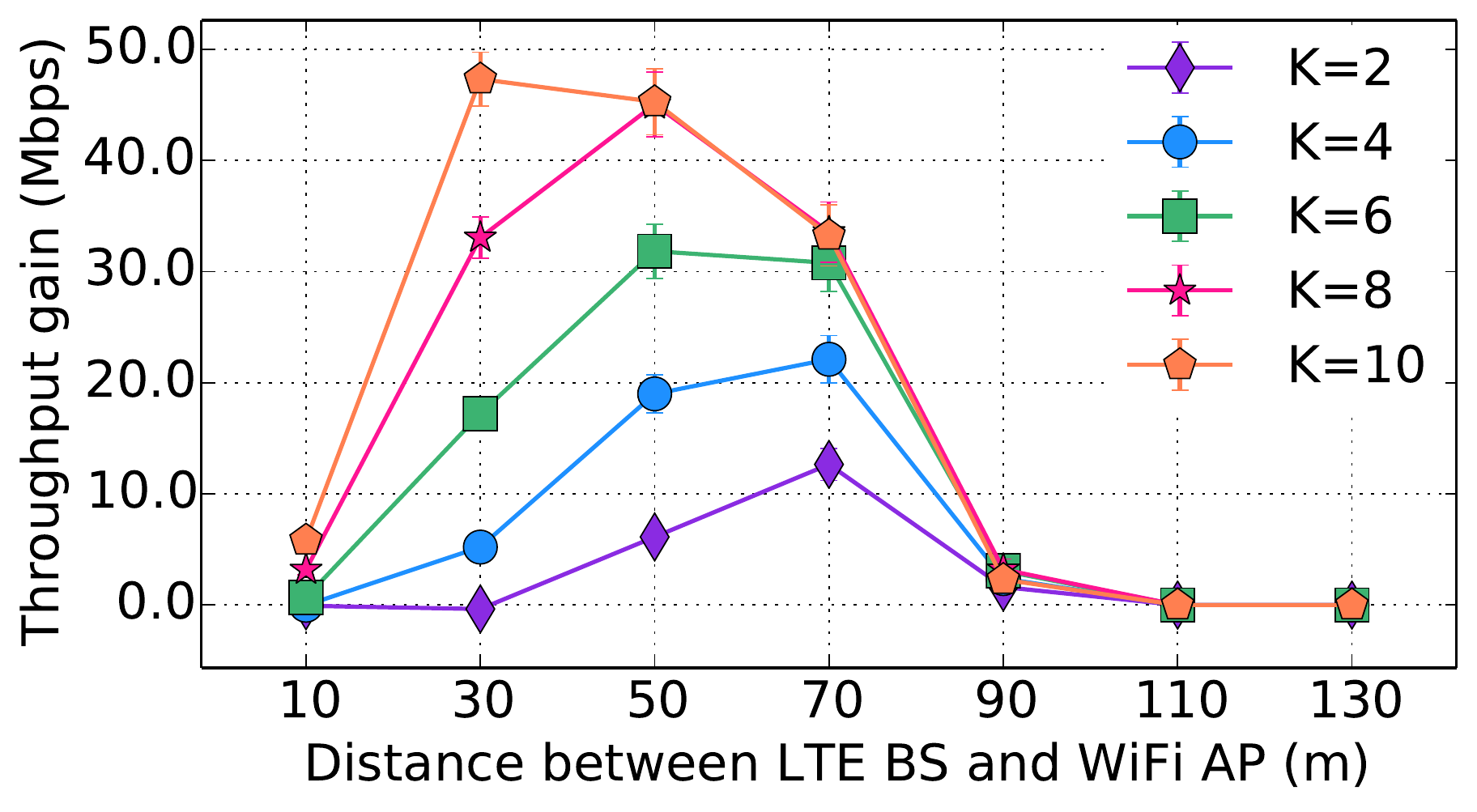}\label{fig:eval_impact_antenna_LTE}} \hfill
  	\subfloat[WiFi throughput gain.]{\includegraphics[width=\linewidth]{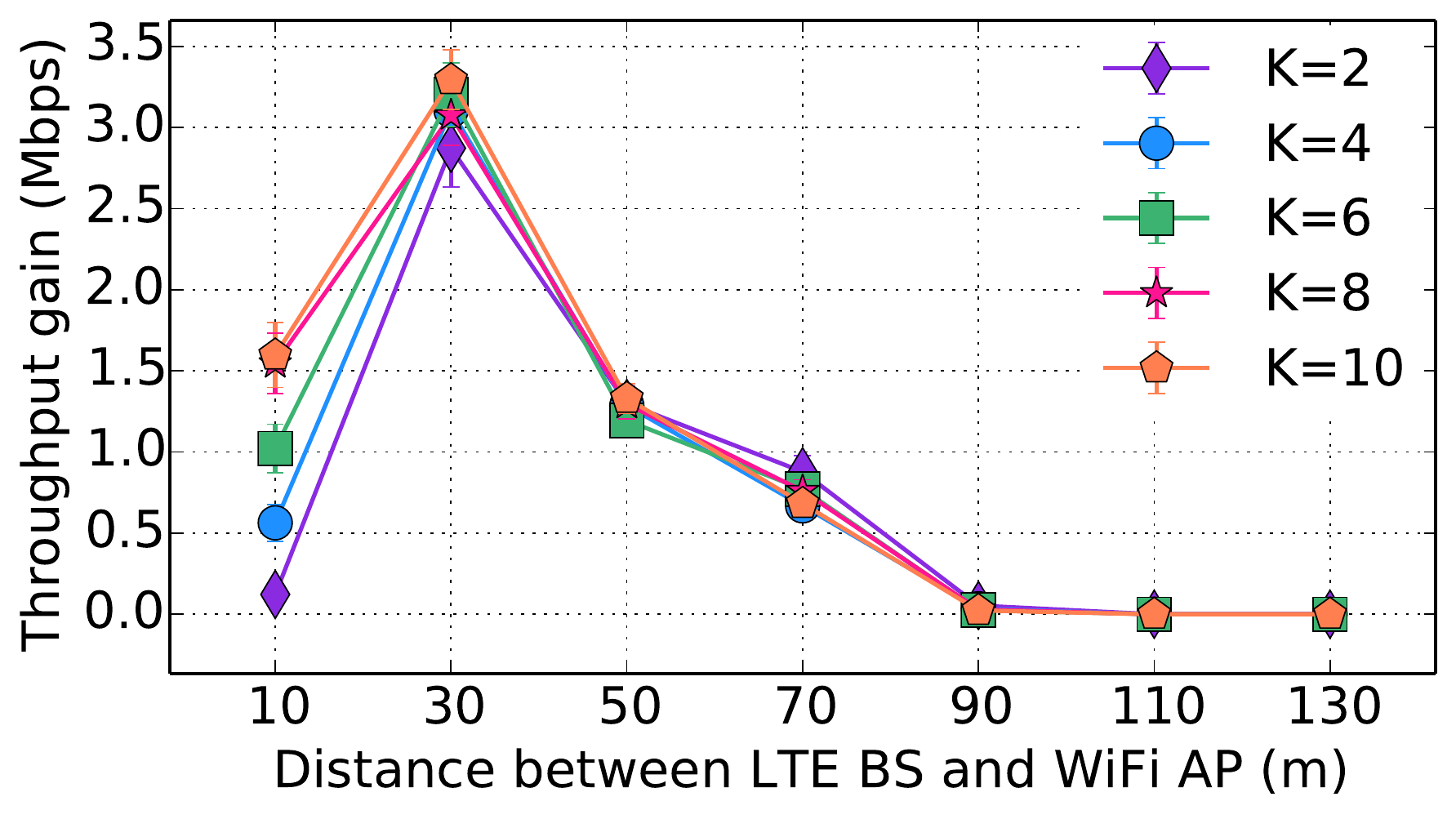}\label{fig:eval_impact_antenna_WiFi}} 
	\caption{Change in the throughput gain over \nonull~ with increasing LTE-U and WiFi separation distance under various LTE-U BS antenna settings, $\numUsersWiFi=8$.\label{fig:eval_impact_antenna}}
\endminipage\hfill
\minipage{0.33\textwidth}
\subfloat[Airtime.]{\includegraphics[width=\linewidth]{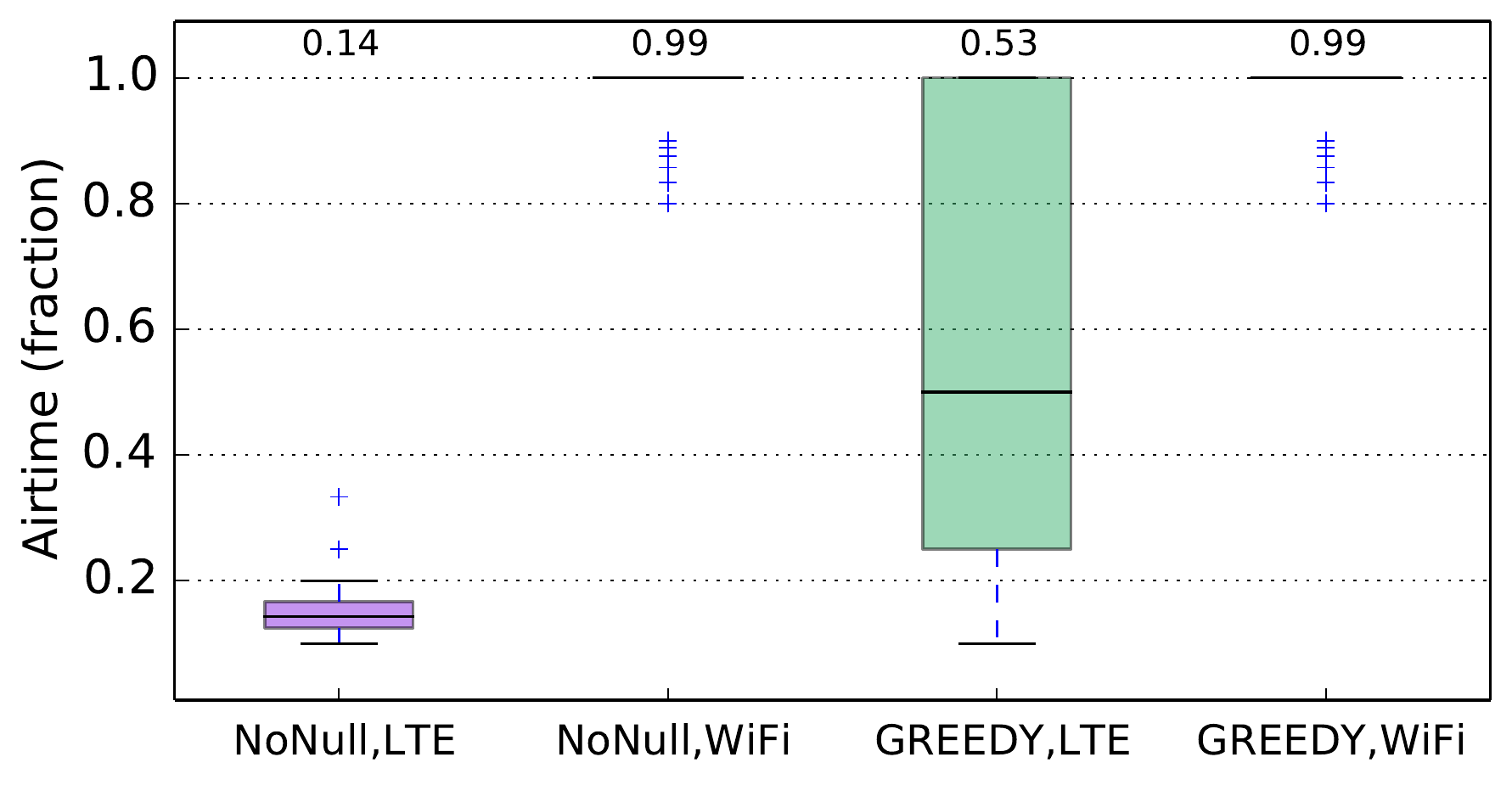}\label{fig:box_plot_airtime}} \hfill
			\subfloat[SNR in dB.]{\includegraphics[width=\linewidth]{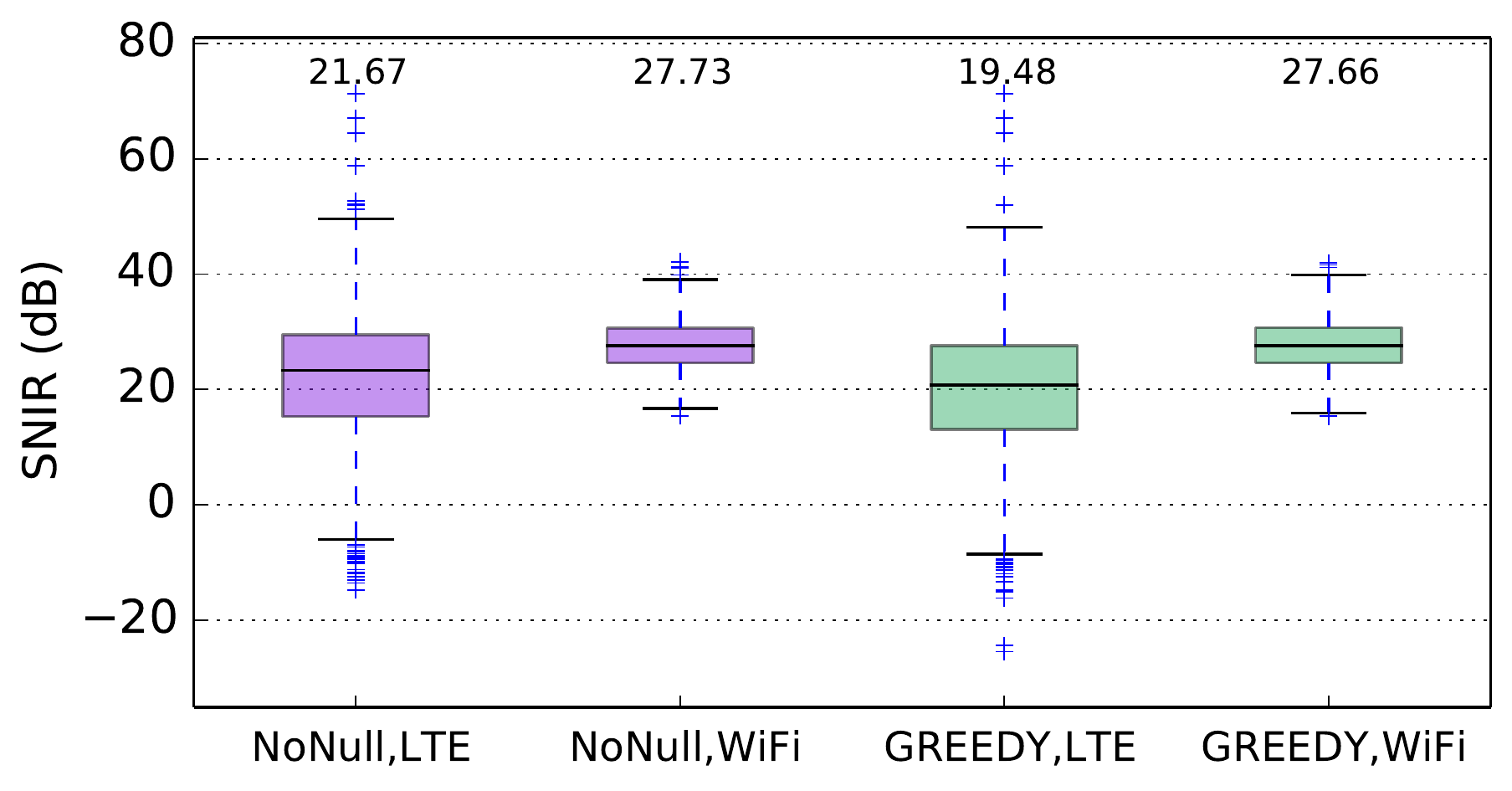}\label{fig:box_plot_snr}} 
\vspace{0.2cm}\caption{Airtime and average SNR under \nonull~ and \heuristicname~ for $\distanceWiFiLTE=30$ m and $\numLTEantenna=10$. Number above each box represents the mean value.\label{fig:box_plots_d30}}
\endminipage\hfill
\minipage{0.32\textwidth}%
	\subfloat[LTE-U throughput gain.]{\includegraphics[width=\linewidth]{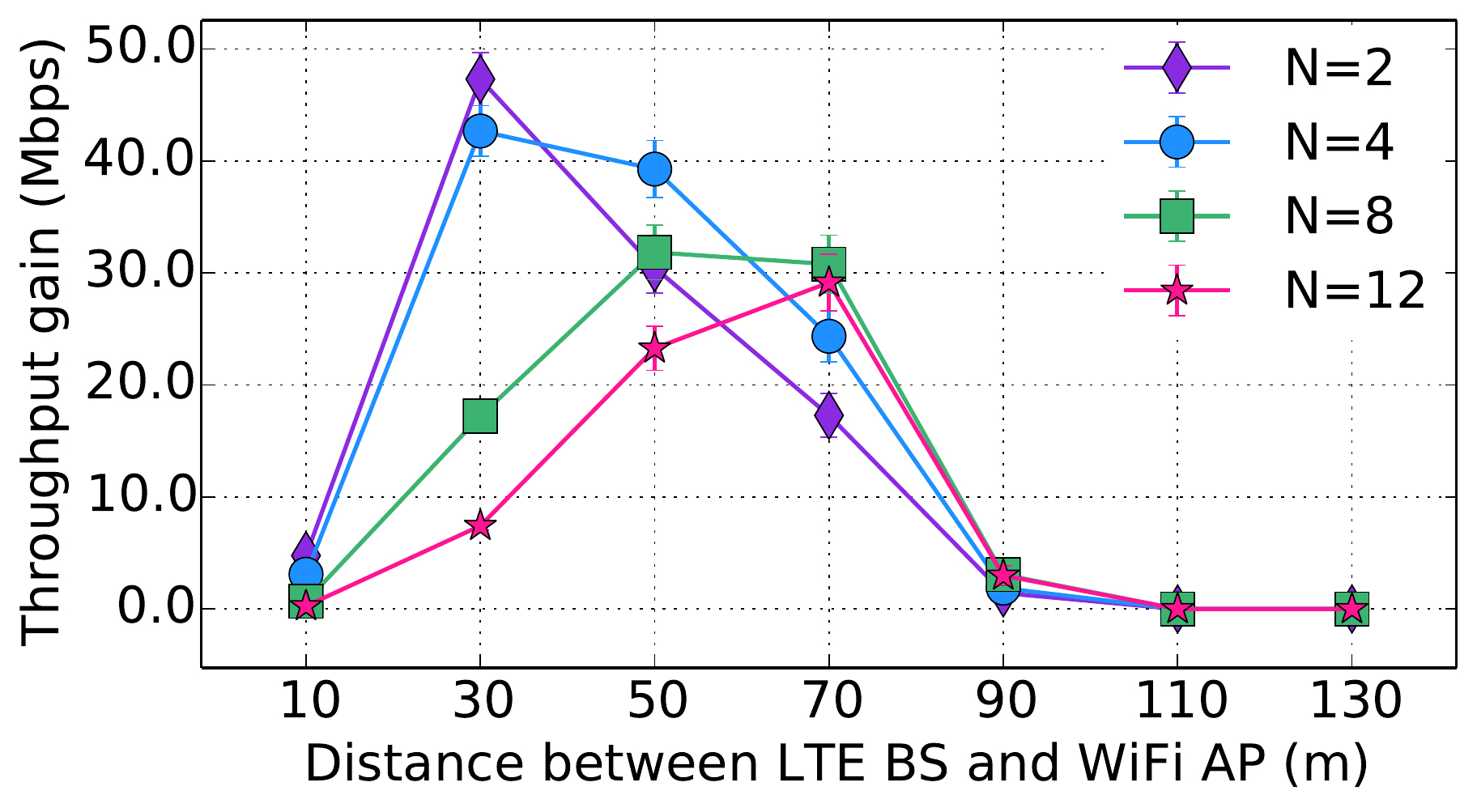}\label{fig:eval_impact_users_LTE}} \hfill
		\subfloat[WiFi throughput gain.]{\includegraphics[width=\linewidth]{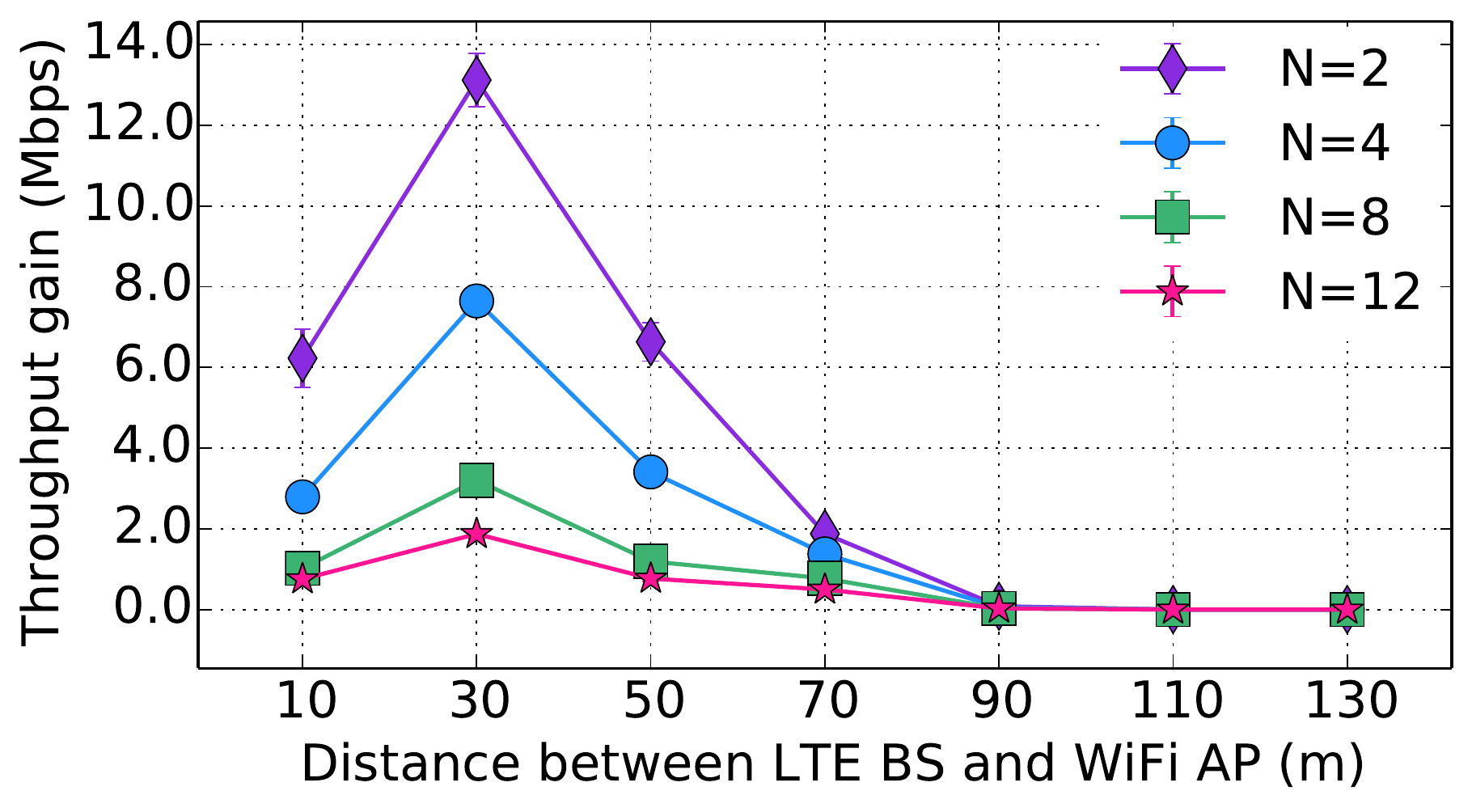}\label{fig:eval_impact_users_WiFi}} 
		\caption{Change in the throughput gain with increasing LTE-U and WiFi separation distance under various number of WiFi stations, $\numLTEantenna=6$. \label{fig:eval_impact_users}}
\endminipage
\end{figure*}

From WiFi's perspective, we observe a similar trend in Fig.~\ref{fig:eval_impact_antenna_WiFi}.
WiFi has throughput gain in all cases for $\distanceWiFiLTE$<90\,m but the gain is markedly lower compared to the LTE-U's gain.
In Fig.~\ref{fig:box_plots_d30}, we show for $\distanceWiFiLTE$=30 m the airtime and SNR under \nonull~ and \heuristicname~ for both LTE and WiFi.
In the figure, we see that the airtime increase in LTE is very significant whereas there is also some decrease in the average SNR due to the loss in antenna diversity.
On the contrary, WiFi experiences almost no change in its SNR and airtime.

\subsection{Impact of number of WiFi users}
Fig~\ref{fig:eval_impact_users} shows the throughput gain of \heuristicname~ over \nonull~ with $\numLTEantenna$=6 antennas at the LTE-U BS for various number of users $\numUsersWiFi$ and under increasing separation distance $\distanceWiFiLTE$. %
As Table~\ref{tab:airtime} shows, higher number of WiFi nodes means lower airtime for the LTE-U cell.
This could also be interpreted as more degrees of freedom or more coexistence gaps in space for the nulling algorithm to exploit.

Regarding LTE-U cell, for short $\distanceWiFiLTE$, Fig.~\ref{fig:eval_impact_users_LTE} shows that nulling brings higher throughput gain for lower number of users. 
In this region, WiFi AP senses the LTE-U BS.
The only way to offer performance improvement also to the WiFi is to null the WiFi AP.
However, WiFi stations, especially the ones in the near proximity of the LTE-U BS, must also be nulled to facilitate interference-free DL traffic at these stations.
If LTE-U BS has enough antennas to null all the nearby stations, the WiFi network will boost its throughput as if there is no coexisting LTE-U network~(as observed in Fig.\ref{fig:eval_impact_users_WiFi}).
Otherwise, i.e., case of many WiFi users, LTE-U may prefer putting coexistence gaps only in the time domain.
Our analysis on average number of nulled stations and AP~(plotted in Fig.\ref{fig:num_nulled_nodes}) show that nulling the AP is preferred only very rarely under higher $\numUsersWiFi$ and short $\distanceWiFiLTE$.

On the other hand, with increasing $\distanceWiFiLTE$, we observe that the highest gain for LTE-U is achieved under higher $\numUsersWiFi$.
For low $\numUsersWiFi$ and high $\distanceWiFiLTE$, these few users might already be far away from the LTE-U BS and there is a lower probability of interference with these stations. 
For higher $\numUsersWiFi$, the expected number of WiFi nodes in LTE-U's ED range is higher, resulting in a need for null steering.

Generally speaking, highest gain for WiFi is achieved when there is a few stations only. These stations will be receiving interference-free traffic mostly when LTE-U cell has sufficient antennas to null them.
As we observe in Fig.\ref{fig:eval_impact_users_WiFi}, WiFi also has non-negative throughput gain under all cases, which proves our claim that our proposal is beyond coexistence; it provides benefits for the LTE-U and WiFi networks.
Considering both Fig.~\ref{fig:eval_impact_antenna} and Fig.~\ref{fig:eval_impact_users}, our experiments suggest that interference nulling provides the highest gains to both networks when their separation distance is moderate, e.g., distances where one network may be hidden to the other.

\subsection{Discussions}\label{sec:discussions}

We have not provided any analysis on the overhead of cooperation in our proposal. 
However, for quasi-static settings, e.g., indoor scenarios where WiFi users 
have mostly very low mobility, we believe that the entailed overhead will be very low.
Moreover, LTE-U BS can implement some schemes to decrease this overhead, e.g., cache the WiFi information and estimate the locations of the nodes if provided some statistics, e.g., about node mobility.

Finally, we considered single user MIMO to be used at the LTE-U BS. 
In case of multi-user MIMO, the sacrificed antenna diversity will be higher leading to a different trade-off function between airtime and best number of nulls.
We leave this aspect to a future work.

%% As always
%!TEX root = ../main.tex

%%%
\section{Related Work}

We can classify the related work on noncoordinated coexistence solutions into two categories depending on where the coexistence solution is implemented: the LTE-U network and the WiFi network. 
Below, we provide an overview of the approaches falling into these two categories with a note that the approaches in the second category are only a few.

\smallskip

\noindent\textbf{Interference management in the LTE-U network:}
LTE-U can manage its interference on neighboring WiFi networks by creating coexistence gaps in several domains: frequency, time, and space. 

\smallskip

\noindent\textit{Coexistence gaps in time:}
A simple coexistence scheme reuses the concept of almost blank subframes and subframe puncturing in LTE-U in order to create coexistence gaps in time domain~\cite{beluri2012mechanisms,almeida2013enabling,zhang2015coexistence}. Works adapting the LTE-U's duty-cycle length all fall into this category.

\smallskip

\noindent\textit{Coexistence gaps in frequency:}
Similar to other spectrum sharing scenarios, frequency-domain sharing is the first step in coexistence of LTE-U and WiFi.
An LTE-U BS seeks for a clear channel to avoid impairing incumbent WiFi networks.
In \cite{al20155g}, a co-existence scheme is proposed that deals with the available channels of the unlicensed band as one pool. This means that the LTE-U will switch between various channels all the time to avoid the excessive use of one channel resulting in coexistence gaps in frequency domain.

\smallskip

\noindent\textit{Coexistence gaps in space:}
Coexistence can be achieved in space domain, e.g., changing the transmission power to adapt the interference region. 
Chaves et al.~\cite{chaves2013lte} proposed an LTE UL power control with an interference-aware power operating point which represents an alternative to the time-sharing approach for LTE-U/Wi-Fi coexistence.  
By a controlled decrease of LTE-U UEs’ transmit powers, the interference caused to neighboring Wi-Fi nodes diminishes, thus creating WiFi transmission opportunities as WiFi nodes detect the channel as vacant.
Our work falls into this category as we also create interference-free spaces in the WiFi cell. 
However, our proposal differs from existing works in many ways, e.g., it is coordinated coexistence exploiting the antenna resources of LTE-U BSs to achieve both gains at the LTE-U and the WiFi.

\smallskip
\noindent\textit{Approaches which aim at increasing LTE-U airtime}: There are also some approaches which apply a mixture of solutions with a main goal to increase LTE-U's airtime. 
Power control is one way to decrease the interference range of the LTE-U BS and in return increase its duty cycle. 
Another approach proposed in \cite{chen2016rethinking} is to  handover some of the WiFi users to the LTE-U cell so that LTE can gain some airtime by effectively using its spectral capacity to satisfy the transferred users' traffic requirements. 

\smallskip

\noindent\textbf{Interference management in the WiFi network:}
Although majority of the literature focuses on the LTE-U side, WiFi can also be equipped with mechanisms to be aware of neighboring LTE-U networks and strategize accordingly, e.g., move to another channel.
The only work in this category is WiPLUS introduced by Olbrich et al.~\cite{olbrich2017wiplus}.
WiPLUS is a noncoordinated coexistence solution where interference mitigation is performed solely by the WiFi network and hence being transparent to LTE-U. 
The proposed WiPLUS ranges from simple approaches where WiFi in order to mitigate interference towards LTE-U is simply abandoning the affected channel to complex interference-aware medium access and channel bonding where WiFi adapts its PHY/MAC parameters so that its transmissions are not colliding with scheduled and hence predicted LTE-U transmissions.

%% Conclusions
%!TEX root = ../main.tex

%%%
\section{Conclusions \& Future Work}\label{sec:conc}

It is crucial that operation of LTE networks in the unlicensed spectrum does not jeopardize WiFi, which is very coexistence-friendly owing to its listen-before-talk medium access nature.
We have proposed a coordinated coexistence scheme for WiFi and LTE-U networks where LTE-U BSs equipped with multiple antennas create coexistence gaps in space domain by means of cross-technology interference nulling towards co-located WiFi nodes in the interference range.
We provided algorithms to compute the WiFi nodes to be nulled.
Simulation results reveal that proposed cooperation offers benefits to both LTE-U and WiFi in terms of improved capacity and faster channel access.
The proposed cooperative scheme can be implemented on top of a cross technology communication channel like LtFi~\cite{gawlowicz2017ltfi}.
For future work, we plan to implement a prototype using SDR platform which would allow us to analyze the performance gain under more realistic assumptions, i.e., imperfections in the nulling/beamforming process and mobility.
Another direction worth exploring is the LTE-U in the UL.

%% This section will be only in the long technical report version 

\section*{Acknowledgment}
This work has been supported by the European Union's Horizon 2020 research and innovation program under grant agreement No 645274~(WiSHFUL project).
\bibliographystyle{ACM-Reference-Format}
\bibliography{biblio}
%!TEX root = main.tex
% This section will be in the longer version, technical report

\section{Appendix}\label{sec:appendix}
Fig.~\ref{fig:num_nulled_nodes} plots the average number of nulled WiFi stations and the AP.
\begin{figure}[!th]
\subfloat[Various number of LTE-U BS antennas.]{\includegraphics[width=0.6\linewidth]{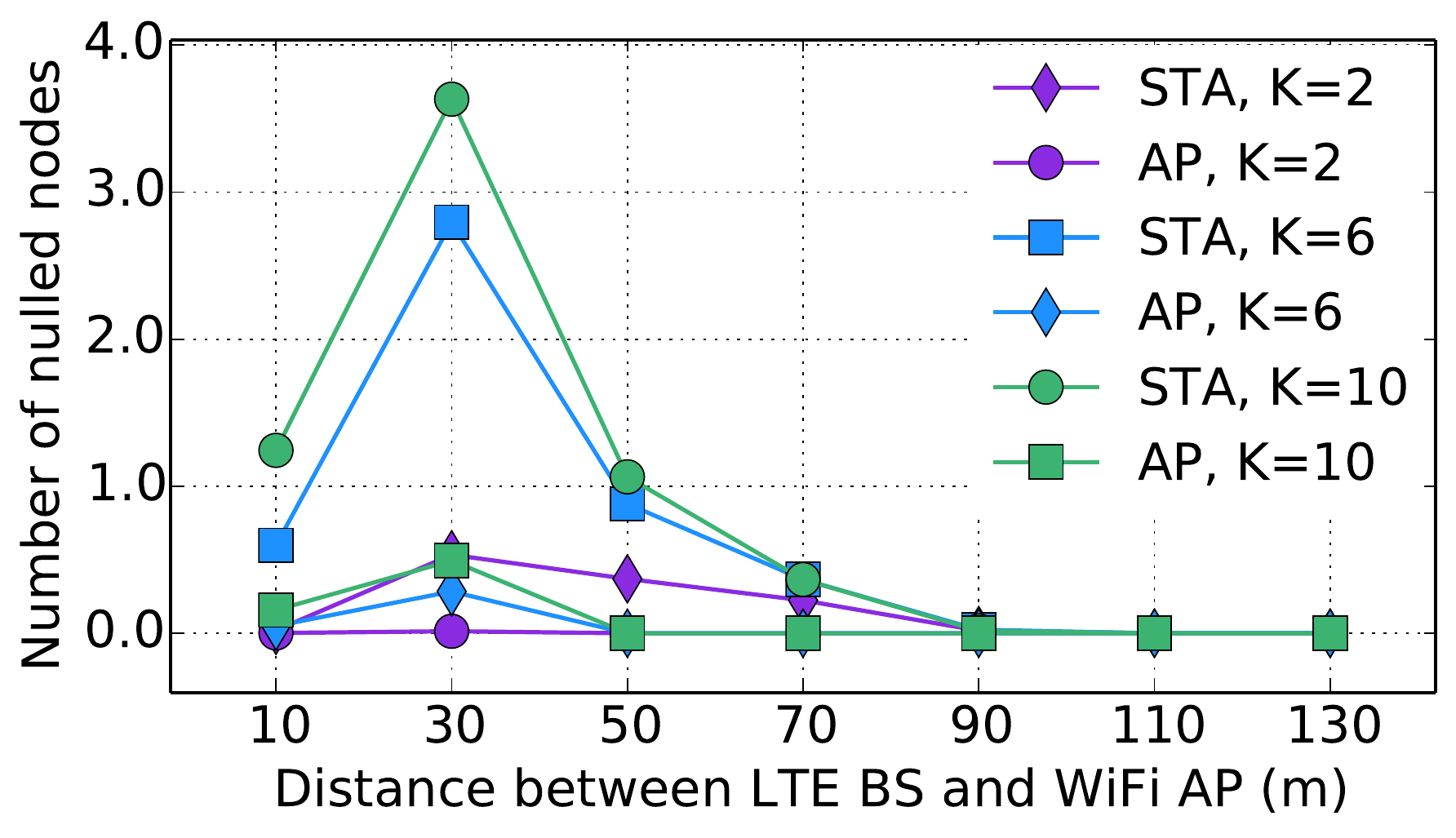}\label{fig:eval_impact_antennas_num_nulled}}\\
\subfloat[Various WiFi stations.]{\includegraphics[width=0.6\linewidth]{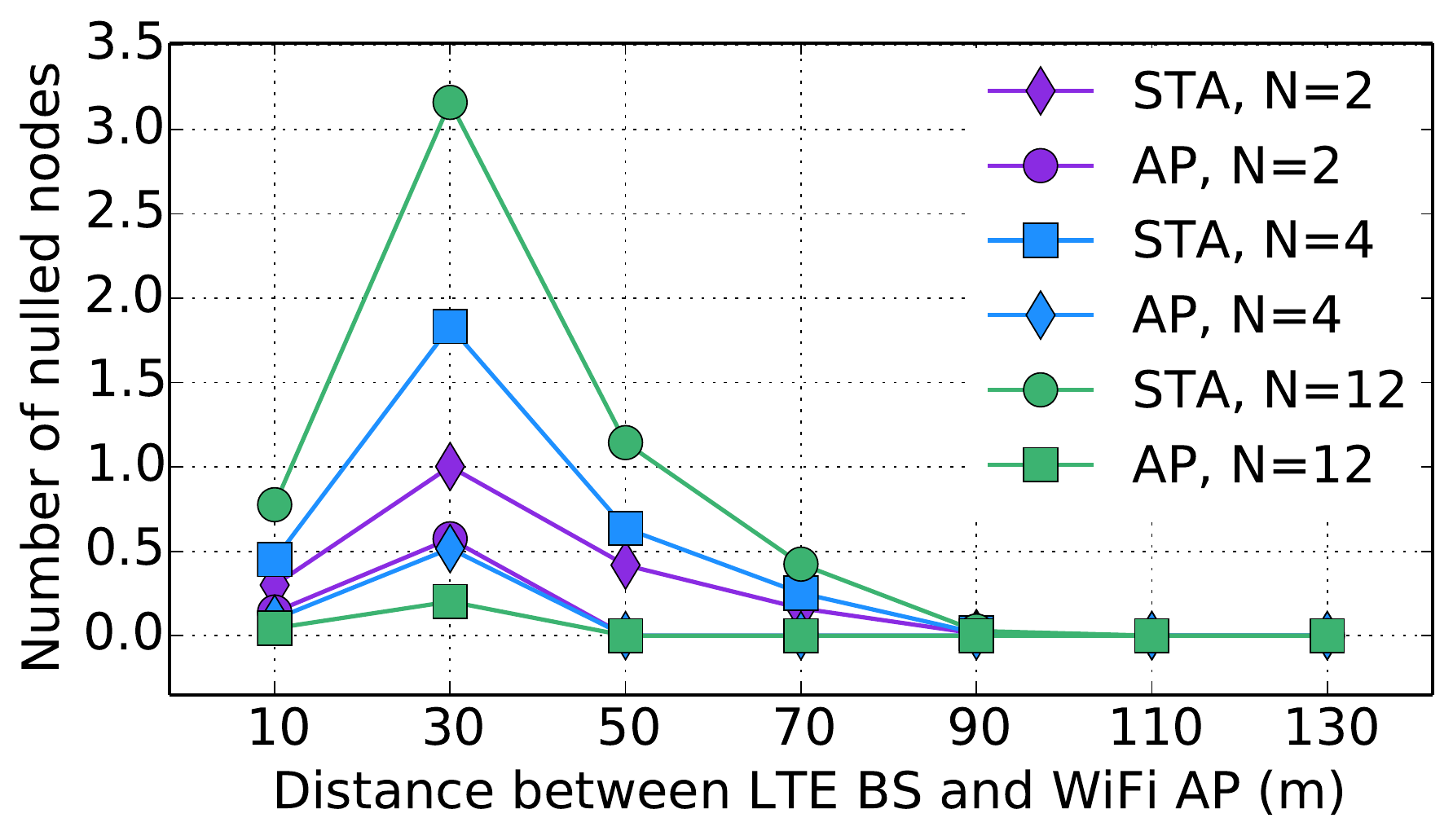}\label{fig:eval_impact_users_num_nulled}}
\caption{Number of nulled stations and the AP with increasing distance.}\label{fig:num_nulled_nodes}
\end{figure}
\end{document}